\documentclass{article}

\usepackage[utf8]{inputenc}
\usepackage{xcolor}
\usepackage{graphicx}
\usepackage{amsmath}
\usepackage{amsfonts}
\usepackage{gensymb}
\usepackage{cite}
\usepackage{authblk}

\newcommand{\bra}[1] {\left<#1\right|} 
\newcommand{\ket}[1] {\left|#1\right>} 
\newcommand{\expect}[2][] { 
  \if\relax\detokenize{#1}\relax
    \left<#2 | #2\right>
  \else
    \if\relax\detokenize{#2}\relax
      \left< #1 \right>
    \else
      \bra{#2} #1 \ket{#2}
    \fi
  \fi
}

\makeatletter
\newcounter{savesection}
\newcounter{apdxsection}
\renewcommand\appendix{\par
  \setcounter{savesection}{\value{section}}%
  \setcounter{section}{\value{apdxsection}}%
  \setcounter{subsection}{0}%
  \gdef\thesection{\@Alph\c@section}}
\newcommand\unappendix{\par
  \setcounter{apdxsection}{\value{section}}%
  \setcounter{section}{\value{savesection}}%
  \setcounter{subsection}{0}%
  \gdef\thesection{\@arabic\c@section}}
\makeatother

\newenvironment{dict}[1]{
    \noindent \underline{\textbf{#1}}\\
}
{
\\\\
}

\title{A Conformal Field Theory Primer in $D\geq3$}
\author[1]{Andrew M. Evans}
\author[2]{Alexandra Miller}
\author[3]{Aaron Russell}
\affil[1,2,3]{Department of Physics and Astronomy, Sonoma State University, Rohnert Park, CA 94928}
\affil[1]{Department of Physics and Astronomy, UC Irvine, Irvine, CA 92697}
\date{(Dated: September 18, 2023)}

\usepackage{hyperref}

\begin{document}

\maketitle
\begin{abstract}
    This primer is an introduction to Conformal Field Theory in $D\geq3$. It is designed to introduce the reader to many of the important foundational concepts and methods in CFT. In it, big picture ideas are prioritized over technical details, which can be found in many of the other excellent reviews that already exist. It is written primarily with upper division undergraduate Physics majors, master’s students, and PhD students in mind. However, it should additionally be useful to anyone who is interested in learning the basics of CFT and has some foundation in undergraduate Physics coursework.
\end{abstract}
\newpage
\tableofcontents
\newpage

\section{Preliminaries}
\subsection{Who is this primer for and How to use it?}
This primer is designed to introduce upper-division undergraduate Physics majors, graduate students, and anyone else who is interested to Conformal Field Theory (CFT). After reading it, one can expect to be familiar with much of the terminology, the conceptual foundation, and some of the mathematics. It is not designed to be a technical introduction to the topic, as there are many excellent technical introductions already in existence (see, e.g. \cite{bible,simmonsduffin2016tasi,Rychkov_2017,pirsa_PIRSA:11110050,qualls2016lectures}). Many of these introductions are written with advanced graduate students in mind. This primer, however, is written in such a way as to be accessible to upper division undergraduates and all graduate students in Physics or Astrophysics. It will give them enough background to start diving into the literature, including setting them up to better understand the aforementioned technical reviews.

CFT builds on many areas of physics. Therefore, in this Preliminary Section, we review many topics that will be helpful in setting the foundation for CFT. The reviews in this section are in no way intended to replace a full course on the topics, but are designed to remind the reader of some key aspects that will be helpful in building up to CFT. As such, some more advanced readers may decide to simply skip parts of this section or go straight on to section \ref{intro}. Beyond many standard undergraduate topics, these reviews include a discussion of Relativity and Quantum Field Theory (QFT), which many will not see until graduate school. It is highly recommended that readers unfamiliar with QFT read through this piece.

For anyone reading this primer, it will be helpful if you have already completed the following undergraduate level courses:
\begin{itemize}
\item Multivariable Calculus
\item Linear Algebra and Differential Equations
\item An upper division Modern Physics course
\item Mathematical Methods for Physics
\item Analytic Mechanics (Lagrangian Mechanics)
\item At least one semester of Quantum Mechanics (ideally, a spins first approach, such as that presented in Townsend \cite{townsend} or McIntyre \cite{mcintyre})
\end{itemize}

Please note that Appendix \ref{dictionary} consists of a Quick Reference Dictionary. This dictionary includes brief descriptions of key terms in CFT, which the reader may find useful.

\subsection{Lagrangian Mechanics} \label{classical}

Ideally, the reader of this primer will already have some familiarity with Lagrangian Mechanics from an upper division Mechanics course. Here, we give a brief reminder of the key points from Lagrangian Mechanics that will be helpful in building up to Conformal Field Theory (CFT). Specifically, the reader needs to be familiar with the idea of a \textit{Lagrangian}.

In lower division physics courses, we are taught that forces act on systems and it is these forces that cause objects to move in the way they do, $F=ma$. However, this is not the only way we can think about these classical systems. Instead of starting with forces, we can define our system with one function\footnote{Technically, the Lagrangian is a functional, which differs from a function in that it takes a function as its input, rather than a variable.}, called a Lagrangian. In many cases, this function is simply the kinetic energy minus the potential energy $L=T-V$. So, we are focusing on energy rather than on forces. In fact, in Lagrangian Mechanics you do not deal with forces at all. Once you are given (or determine) the Lagrangian for a system, you can compute the equations of motion and ultimately solve for the dynamics of a system, given a set of initial conditions. You do this by considering the \textit{action}, $S$, which is given by the integral over the Lagrangian, $S=\int dx L$. Lagrangian Mechanics is based on a the following fundamental principle: \textbf{Classical systems will follow the trajectory that extremizes the action}.

In Quantum Field Theory, which is discussed in section \ref{QFT}, the Lagrangian is often taken as the starting point. It includes all of the information about how the different fields will interact with each other and from it we compute things like scattering amplitudes in Particle Physics. We will learn that (quantum) Conformal Field Theories are just special cases of QFTs with additional symmetry. This additional symmetry will allow us to think about CFTs in a very different way from QFTs and often we will not concern ourselves with the Lagrangian. Indeed, for some CFTs, the Lagrangian is not known and it might not be possible to even find a Lagrangian description.

\subsection{Quantum Mechanics} \label{QM}
One can consider both classical and quantum Conformal Field Theories. In this primer, we will primarily be concerned with quantum CFTs. Therefore, it is important to review a bit of Quantum Mechanics (QM). Ideally, the reader has already had some experience with undergraduate QM. It will be especially helpful if the student has taken a class with a Spins First Approach (such as is presented in \cite{townsend} and \cite{mcintyre}), as opposed to a class that solely focuses on wave-functions. 

Let's start by talking about some major conceptual differences between Classical and Quantum Mechanics. In the classical view of physics, the state of any system is definite. This means that any identical measurements made on identical systems will yield identical results. For example, consider a ball experiencing projectile motion. If this experiment was performed several times with the same initial conditions (i.e. same velocity and launch angle) and all external conditions held constant, then at time $t$ a measurement of the $x$-position, $y$-position, and velocity would yield the exact same result for every trial. 

Alternatively, the quantum mechanical view of physics tells us that we live in a probabilistic universe. To illustrate this, consider measuring the position of a particle at time $t$ moving through free space. If this experiment is repeated several times on identically prepared particles (i.e. particles in the same state, $|\psi\rangle$), then the position measured may differ for each trial. 

Even though we generally can’t give an exact location of the particle without making a measurement, we would still like to be able to talk about the particle’s position in some fashion, so we introduce the expectation value. In the experiment above, if you were to take the average of all of the measured positions of the particle, this average would be very close to the expectation value. We say “close” and not “exactly” because the expectation value is actually what you would get if you took the average measurement over an infinite number of trials, which of course is impossible to do in reality.

As with position, all observable quantities (e.g. momentum, energy)  have this probabilistic nature in QM and therefore all observables can be talked about in terms of their expectation values. Then arises the question of how to calculate these expectation values. In order to do this we will introduce the idea of an operator, $\hat{\mathcal{O}}$.

At the most fundamental level, an operator is an instruction to transform whatever is immediately to the right of the operator. To start with a familiar example, consider the equation $\frac{d}{dx}x^2=2x$. Here we have the differential operator $\frac{d}{dx}$ telling us to transform $x^2$ into its derivative, $2x$. In QM, instead of transforming scalars, operators transform vectors and take the form $\hat{\mathcal{O}}|\psi\rangle$ where $\hat{\mathcal{O}}$ is an operator and $|\psi\rangle$ is a vector that represents the state of the system.

Linear Algebra forms the mathematical foundation of QM. Here, we will review all of the important types of mathematical objects:
\begin{itemize}
    \item Vectors $|\Psi\rangle$ describe states of the system.
    \item Basis Vectors $|n\rangle$ can be used to build general states of your system. If you have a complete basis, any general state can be represented as a superposition over the basis states, i.e. $|\psi\rangle=\sum_n c_n|n\rangle$.
    \item Dual Vectors $\langle \Psi |$ live in their own vector space and allow us to take inner products. Here, $\langle\Psi |=(|\Psi\rangle)^\dagger$ and the dagger signifies the Hermitian conjugate.
    \item Operators $\hat{\mathcal{O}}$ act on vectors to transform them into new vectors. In general, $$\hat{\mathcal{O}}|\Psi \rangle=|\Psi '\rangle$$ $$\langle \Psi |\hat{\mathcal{O}}^\dagger=\langle \Psi '|$$
    \item Hermitian Operators $\hat{\mathcal{O}}^\dagger=\hat{\mathcal{O}}$ represent observables, such as momentum, energy, position, etc.
    \item Unitary Operators $\hat{\mathcal{O}}^\dagger\hat{\mathcal{O}}=\hat I$ preserve the normalization of states and therefore are used to represent time-evolution of the system.
    \item Inner Products $\langle \phi |\psi \rangle$ tell you the overlap between the states $| \phi \rangle$ and $| \psi \rangle$ (roughly, how similar they are).
    \item Expectation Values $\langle \hat{\mathcal{O}} \rangle=\langle \Psi |\hat{\mathcal{O}}|\Psi \rangle$ give you the average value of a particular observable after a series of measurements on identical states.
    \item Eigenvalues $\lambda$ and Eigenvectors $|\lambda\rangle$ of an operator $\hat{\mathcal{O}}$ are given by $\hat{\mathcal{O}}|\lambda\rangle=\lambda |\lambda\rangle$. Eigenstates of Hermitian operators can be used to form a basis for your vector space.
    \item Commutators $[\hat{\mathcal{O}}_1,\hat{\mathcal{O}}_2]=\hat{\mathcal{O}}_1\hat{\mathcal{O}}_2-\hat{\mathcal{O}}_2\hat{\mathcal{O}}_1$ tell you about how acting with operators in a different order may (or may not) change the result. One important fact is that two operators that fail to commute cannot have simultaneous eigenstates.
\end{itemize}
With these in mind, we can present the Dirac Von Neumann axioms of QM:
\begin{itemize}
    \item[] \textbf{Axiom 1.} States are represented by vectors in a Hilbert space, $|\Psi \rangle$.
    \item[] \textbf{Axiom 2.} Observables are associated with Hermitian operators, $\hat{\mathcal{O}}^\dagger=\hat{\mathcal{O}}$.
    \item[] \textbf{Axiom 3.} Expectation values are given by $\expect[\hat{\mathcal O}]{\psi}$.
\end{itemize}

These form a foundation for Quantum Mechanics\footnote{There is ongoing current research in the field of Quantum Foundations that questions what the best list of axioms are or even whether thinking about forming the foundation from a list of axioms is the best approach.}. As we will be dealing with quantum CFTs in this primer, these axioms will be taken to be true. Additionally, quantum theories are generally taken to be unitary, which means that their dynamics are governed by a unitary operator. This is often listed as its own axiom. In order for the probabilistic interpretation of QM to make sense, states of our system must be properly normalized $\langle\psi|\psi\rangle=1$. This guarantees the total probability for all possible outcomes of a given measurement is one. If we consider the evolution of the state of our system, then we want this normalization condition to hold. That is guaranteed if the time-evolution operator is given by a unitary operator. In CFT, we can learn a lot by enforcing the condition of unitarity. In particular, we will see that it bounds the allowed spectrum of operators in our theory, as is discussed in section \ref{unitaritybound}.

\subsection{Symmetry and Group Theory} \label{group}

An understanding of symmetries is imperative to our understanding of CFTs. Therefore, in this section, we will start by discussing symmetries in general. We will then discuss what it means for a physical system to have certain symmetries, and finally we will introduce a bit of Group Theory, which forms a mathematical foundation for thinking about symmetries.

A symmetry of an object can roughly be thought of as an action which can be preformed that leaves the object in an indistinguishable state from the original configuration. As an example, consider 90 degree rotations of a square.
\begin{center}
    \includegraphics[scale=0.15]{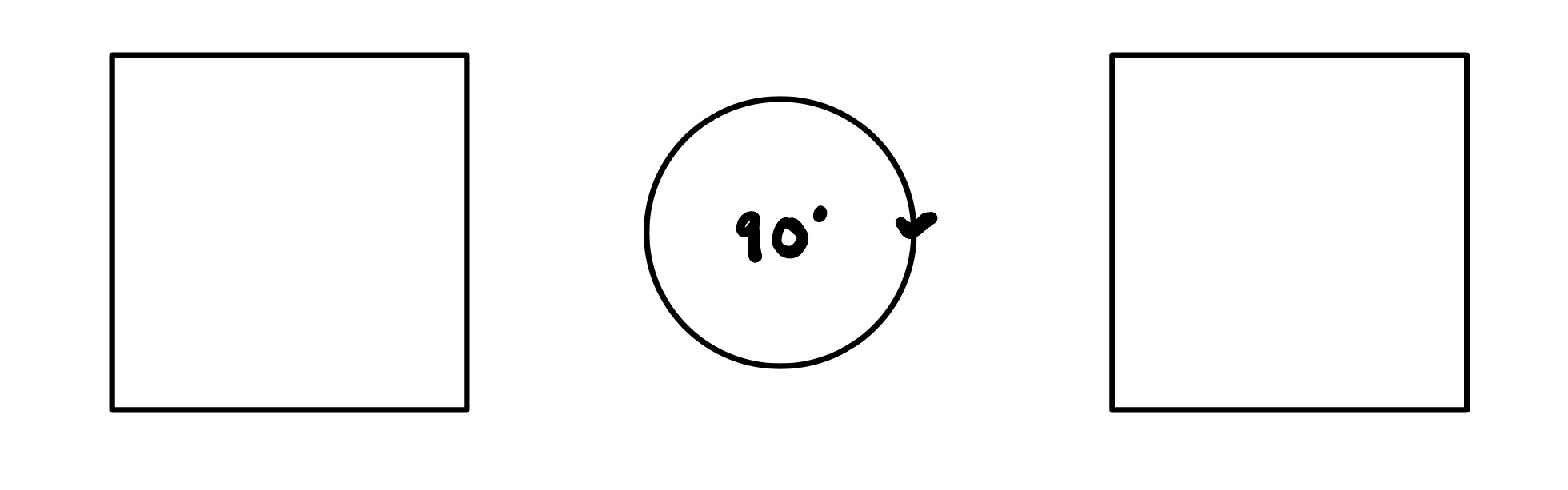}
\end{center}
This figure is a little bit hard to understand, because it looks like nothing has happened (which is a good sign that we have found a symmetry). To remedy this, we will consider the same situation, but this time we will label the corners of the square.
\begin{center}
    \includegraphics[scale=0.15]{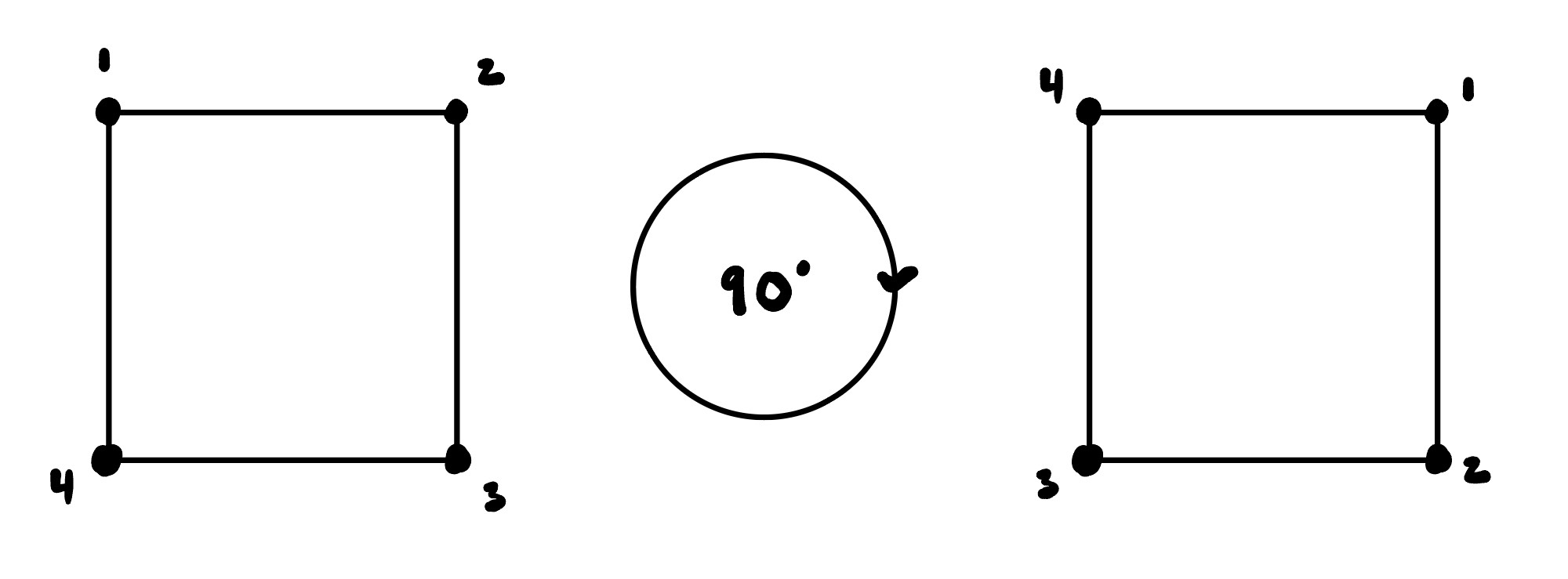}
\end{center}
Now we can see that in fact something did happen! The action of rotating the square 90 degrees appeared to leave the square the same as before, but with labels we can see that something did actually change. This hopefully feels very familiar. But, what does it mean for a physical system to have symmetries? And, given a system, how can we determine what the symmetries are? Like with geometric symmetries, symmetries in physics can be found by searching for transformations that leave the system unchanged in some way. 

There are multiple ways one can imagine for a physical system to have a symmetry. For instance, perhaps the particular trajectory of a system has a certain symmetry (e.g. a projectile in free-fall might have reflection symmetry about the highest point of the path). These symmetries of particular solutions for a physical system will not be the types of symmetries we are primarily interested in. Instead, we will declare that a certain system has a symmetry if the action describing the system has that symmetry. Note that the action is discussed in section \ref{classical}. Specifically, the action, $S$, is given by $$S=\int d^dx\mathcal L$$ where $\mathcal L$ is the Lagrangian of your system. A symmetry is any transformation on your system that preserves the action. So, if you perform a transformation, $T$, on your system such that $\mathcal L \rightarrow \mathcal L'$ and find that $$S'=\int d^dx\mathcal L'=S$$ then $T$ is a symmetry of your system.

\textit{Noether's Theorem} teaches us that for each of these symmetries of the action, there is a conserved charge. For instance, conservation of energy comes from time-translation symmetry, while conservation of momentum comes from spatial-translation symmetry.

As an aside, one could also consider symmetries of the equation of motion itself. One can show that symmetries of the action always imply symmetries of the equation of motion, but not necessarily the other way around. Noether's Theorem only applies to symmetries of the action and when we talk about conformal symmetries, we will be talking about symmetries of the action.

\subsubsection{A Little Group Theory}

To develop the mathematical tools needed to understand symmetry we will need to understand groups (\cite{georgi} is a nice introduction to group theory for physicists). A group is a set of objects $a,,b,c,\dots$ along with a rule for combining the objects (denoted $(\circ)$), such that the following are all true:

\begin{enumerate}
    \item $a\circ b \circ c = (a\circ b) \circ c = a\circ (b \circ c) $ \textbf{(Associativity)}
    \item There exists $e \in G $ such that $e  \circ  a = a \circ e = a $ for all $a \in G$ \textbf{(Identity)}
    \item For each $a  \in G$, there exists $a^{-1}$ such that $a\circ a^{-1}  = a^{-1} \circ a = e$ \textbf{(Inverse)}
    \item For all $a,b \in G$, $a \circ b \in G$ \textbf{(Closure)}
\end{enumerate}
Any set of mathematical objects that follows these rules forms a group! 

Groups can either have a finite set of elements or an infinite set. In Physics, we are especially interested in a specific type of groups known as \textit{Lie Groups}, which are continuous and therefore have an infinite set of elements. To solidify all of this, let's consider two very important Lie Groups in Physics:

\vspace{0.5cm}
\noindent \underline{Orthogonal Group, O(N)}: In this group, the elements are the set of $N\times N$ orthogonal matrices, where orthogonal means $O^T=O^{-1}$. The operation for combining matrices is standard matrix multiplication.

\vspace{0.5cm}
\noindent \underline{Special Orthogonal Group, SO(N)}: This group is the same as the Orthogonal Group, but it has the additional constraint that the set only includes orthogonal matrices with determinant equalling +1.

\vspace{0.5cm}
You might be wondering what all of this has to do with symmetry. The elements of these groups will correspond to symmetry transformations of our system. For instance, $SO(3)$ corresponds to the set of rotation matrices in 3D space. For that reason, this is often referred to as the \textit{Rotation Group}. So, if you declare a system is symmetric under $SO(3)$, that means it has rotational symmetry in three dimensions.

\subsubsection{Lie Algebras} \label{lie}
When we said Lie Groups are continuous, what we meant was that their elements depend smoothly on a continuous set of parameters. If this is true, then we can consider the group elements that are close to the identity. For instance, in the case of the rotation group, these would be the elements that correspond to infinitesimal rotations. These special elements will be referred to as \textit{Generators}. That is because if you start by only considering infinitesimal rotations, then you can build up to (or generate) larger rotations by acting with the Generators many times.

Once the generators of the group are found, one can compute what is known as the \textit{Lie Algebra} of the group. This is going to tell us how much ordering matters when acting with different elements of the group. For instance, it is hopefully familiar that performing a rotation about the $x$-axis, followed by a rotation about the $y$-axis in general leads to different results. The algebra for the rotation group will include this fact. The algebra is computed by determining the commutation relations between all of the generators $G_i$ of the group, e.g $[G_1,G_2]$. For a specific example, see the section on the Poincar\'e Group in section \ref{poincare}. Though, you will need notation introduced in the next section to fully understand that.

\subsection{Special Relativity} \label{SR}

Special Relativity (SR) is the physical description of the universe in the limit where Gravity is so weak that it can be ignored. It is important to use SR when objects are moving very fast (some appreciable percentage of the speed of light). At everyday human speeds, to a very high degree of accuracy, SR will agree with the predictions you would have made in your introductory level Physics class. At higher speeds, effects such as time dilation become increasingly important and the predictions of Galilean Relativity fail.

In General Relativity (GR), we learn that gravity should not be thought of as a force, but instead as a curvature of spacetime itself. We use an object called \textit{the metric} to quantify the geometry of our spacetime. SR is the special case of GR where the metric is flat and therefore there is no gravity. In standard Cartesian ($t,x,y,z,\dots$) coordinates, a flat spacetime is described by the \textit{Minkowski metric} and it is given by\footnote{Please note that there are two common sign conventions for the Minkowski metric, one with mostly plus signs and one with mostly negative signs. This is nothing more than a convention and you must always make note of which is being used. We use primarily plus in this Primer.}
\begin{center}
$ \eta_{\mu \nu} \rightarrow
\begin{bmatrix}
    -1 & 0 & 0 & 0 & \cdots \\
    0 & +1 & 0 & 0 & \cdots \\
    0 & 0 & +1 & 0 & \cdots \\
    0 & 0 & 0 & +1 & \cdots \\
    \vdots & \vdots & \vdots & \vdots & \ddots
\end{bmatrix}$
\end{center}
where this is a $D\times D$ matrix with $D$ being the number of spacetime dimensions. The one negative component is accounting for the difference in the way we treat time. In this primer, we will always be thinking about CFT in flat space. Additionally, we will also often work in \textit{Euclidean Signature}, rather than Lorenzian Signature. This is equivalent to treating time like it is just another spatial dimension. In this case, the metric is just the identity:
\begin{center}
$ \delta_{\mu \nu} \rightarrow
\begin{bmatrix}
    +1 & 0 & 0 & 0 & \cdots \\
    0 & +1 & 0 & 0 & \cdots \\
    0 & 0 & +1 & 0 & \cdots \\
    0 & 0 & 0 & +1 & \cdots \\
    \vdots & \vdots & \vdots & \vdots & \ddots
\end{bmatrix}$
\end{center}
This describes \textit{Euclidean Space}. In introductory level physics, you worked primarily in 3D Euclidean Space. At every instant of time, your system lived in such a space. In SR, we want to treat time and space on equal footing and allow the two to mix in nontrivial ways. This is why we need the 4D Minkowski metric. 

Much of the CFT literature is in Euclidean Space. This is partially because there are some systems where Euclidean signature is appropriate. However, this is also done because it is simpler and, at the end of the day, if one wants to work in Lorenzian signature, one can perform an operation called a \textit{Wick Rotation} to find the correct Lorenzian results. We will not discuss Wick rotation in detail in this Primer. The interested reader may learn more by reviewing David Simmins Duffin's TASI lectures on the topic (add reference).

\subsubsection{Math with Tensors} 
The mathematical foundation of SR involves heavy use of tensors. This will also be true in CFT. In this section, we will review what these are and some mathematical properties of them.

A vector is a rank 1 tensor, which means it is an object with one Lorentz index. These are similar to the vectors you know and love from lower division physics in that we can expand them in terms of their components
  \begin{align*}
    v^\mu &= \begin{bmatrix}
           v_{0} \\
           v_{1} \\
           v_{2} \\
           \vdots \\
           v_{D-1}
         \end{bmatrix}
  \end{align*}
  where the upper index $\mu={0,1,2,\dots D-1}$ is the Lorentz index and it runs over one time dimension and $D-1$ spatial dimensions. For instance, if we are working in four spacetime dimensions, then this object will have four components, representing time plus three spatial components. 

  The thing that makes this a tensor, as opposed to just a regular column matrix, is the way it changes when you make a coordinate transformation. If we start with a vector as represented in some coordinate system, denoted $x^\mu$ and then transform our coordinates (e.g. we rotate our coordinate system) as $x^\mu\rightarrow\tilde{x}^\mu(x)$, then our vector in our new coordinate system is given by
  \begin{equation} \label{vectortransf}
  \tilde{v}^\mu=\frac{\partial\tilde{x}^\mu}{\partial x^\nu}v^\nu
  \end{equation}
  where $\frac{\partial\tilde{x}^\mu}{\partial x^\nu}$ is the Jacobian matrix associated with the transformation. Jacobians are reviewed in Appendix \ref{jacobian}. 

Let's see an example of all of this. Consider a vector in 2D Euclidean space, expressed in Cartesian ($x,y$) coordinates. 
  \begin{align*}
    v^\mu &=  \begin{bmatrix}
           1 \\
           0
         \end{bmatrix}
  \end{align*}
What if we want to perform a coordinate transformation from Cartesian to Polar Coordinates? This transformation is represented by the following equations
  \begin{align*}
    r &= \sqrt{x^2+y^2} \\
    \theta &= \tan^{-1}(y/x)\\
    x&=r\cos(\theta) \\
    y&=r\sin(\theta)
  \end{align*}
To perform the transformation, we need to compute the partial derivatives
  \begin{align*}
    \partial_x r&=\partial_x(\sqrt{x^2+y^2})= \frac{x}{\sqrt{x^2+y^2}}=\cos(\theta) \\
    \partial_y r&=\partial_y(\sqrt{x^2+y^2})= \frac{y}{\sqrt{x^2+y^2}} =\sin(\theta) \\
    \partial_x \theta&=\partial_x(\tan^{-1}(y/x))= \frac{-y}{(x^2+y^2))} = \frac{-\sin(\theta)}{r} \\
    \partial_y \theta&=\partial_y(\tan^{-1}(y/x))= \frac{x}{(x^2+y^2)}=\frac{\cos(\theta)}{r} 
  \end{align*}
Using these in the transformation equation \ref{vectortransf}, we have
 \begin{align*}
    \tilde v^r &= \partial_x r\; v^x + \partial_y r\; v^y = \partial_x r = \cos(\theta)\\
    \tilde v^\theta &= \partial_x \theta v^x + \partial_y \theta  v^y =  \partial_x \theta = \frac{-\sin(\theta)}{r}
  \end{align*}
We can summarize these results as a column vector
  \begin{align*}
    \tilde v^\mu &=  \begin{bmatrix}
           \cos(\theta) \\
           \frac{-\sin(\theta)}{r}
         \end{bmatrix}
  \end{align*}
where the tilde reminds us that this is the representation of the vector in the transformed coordinates.

  In addition to these vectors, which are represented with an upper index, we can also consider objects with a lower index. These go by many names to distinguish them from the vectors with an upper index, including: covectors, 1-forms, covariant vectors, and dual vectors. There is a one-to-one correspondence between our vectors and their associated dual vectors. You compute them by lowering the index using the metric. 
  \begin{equation}
      v_\mu=\sum_{\nu=0}^{D-1}\eta_{\mu\nu}v^\nu=\eta_{\mu\nu}v^\nu
  \end{equation}
  where on the right-hand-side, we have adopted the \textit{Einstein Summation Convention}, which says that any repeated indices should be summed over. We will always use this notation rather that explicitly including the summation symbol in our equations.

  To help make this all explicit, let's consider two vectors $x^\mu$ and $y^\mu$ that represent the spacetime positions of two particles. If we wanted to compute the inner product between these, this is given by $$\eta_{\mu\nu}x^\mu y^\nu=x_\mu y^\mu=-x_0 y_0+x_1 y_1+\dots + x_{D-1}y_{D-1}$$

  Both vectors $x^\mu$ and dual vectors $x_\mu$ are rank 1 tensors because they each have one Lorentz index. We also can consider more general rank $n$ tensors, which will have $n$-indices. In this primer, we will only deal with rank 1 and rank 2 tensors. Like vectors, the defining properties that makes these tensors as opposed to just matrices is the way they transform. If we again consider a coordinate transformation $x^\mu\rightarrow\tilde{x}^\mu(x)$, then a rank two tensor $A^{\mu\nu}$ will transform as
  \begin{equation} \label{tensortransf}
      \tilde{A}^{\mu\nu}=\frac{\partial\tilde{x}^\mu}{\partial x^\rho}\frac{\partial\tilde{x}^\mu}{\partial x^\sigma} A^{\rho\sigma}
  \end{equation}
where again we are using the Einstein Summation Convention. Since there are two repeated indices ($\rho$ and $\sigma$), there are two suppressed sums. We see that we need a Jacobian matrix for each index of the object that is being transformed.

In addition to tensors, we will also need \textit{Tensor Densities}. These are objects that might look like a tensor because they will have Lorentz indices, but they are not technically tensors because they don't transform following \ref{vectortransf} or \ref{tensortransf}. A potentially familiar example of a tensor density is the \textit{Levi-Civita Symbol}, $\epsilon^{\mu\nu\dots}$. This might look like a tensor, but it is technically not a tensor in general geometries. Tensor Densities $D^\mu\nu$ transform as follows
\begin{equation}
    \tilde{D}^{\mu\nu}=\left| \frac{\partial\tilde{x}^\delta}{\partial x^\gamma} \right|^{-\Delta/D}\frac{\partial\tilde{x}^\mu}{\partial x^\rho}\frac{\partial\tilde{x}^\mu}{\partial x^\sigma} D^{\rho\sigma}
\end{equation}
where $\Delta$ is the dimension of $D^\mu\nu$ and $D$ is the spacetime dimension. We will discuss dimensions of objects in section \ref{dimension}.

\subsubsection{The Poincare Group} \label{poincare}

As you can probably guess, flat space has a lot of symmetry. Space will look the same in every direction. You can translate and nothing will change. You can even boost to a reference frame moving at a different velocity (like hopping on a train) and nothing will change. One of the defining properties of Special Relativity is that it is invariant under a set of transformations known as the Poincar\'e Group. In this section, we will tell you everything you need to know about this group and its algebra. This will be a very useful warm-up to the Conformal Group and Conformal Algebra, which we discuss in section \ref{confsymm}.

\textit{The Poincar\'e Group} is the set of transformations that leave the Minkowski metric unchanged. Consider a spacetime with coordinates given by $x^\mu$ and then consider a transformation of these coordinates $$x^\mu\rightarrow\tilde{x}^\mu(x)$$ We can compute the Jacobian associated with this transformation $\frac{\partial\tilde{x}^\mu}{\partial x^\nu}$. The metric is a tensor, so the transformed tensor is given by $$\eta_{\mu\nu}\rightarrow\frac{\partial\tilde{x}^\rho}{\partial x^\mu}\frac{\partial\tilde{x}^\sigma}{\partial x^\nu}\eta_{\rho\sigma}$$
This is said to be a Poincar\'e transformation if
\begin{equation} \label{poincareconstraint}
    \frac{\partial\tilde{x}^\rho}{\partial x^\mu}\frac{\partial\tilde{x}^\sigma}{\partial x^\nu}\eta_{\rho\sigma}=\eta_{\rho\sigma}
\end{equation}

We find a set of differential equations that can be solved in order to find what transformations are allowed. The most general solution to this differential equation is at most linear in $x^\mu$ and is given by $$\tilde{x}^\mu=a^\mu+\Lambda^{\mu}_{\;\nu} x^\nu$$ where $a^\mu$ is a constant vector and $\Lambda^{\mu}_{\;\nu}$ is a constant tensor. Let's plug this solution back into \ref{poincareconstraint} to see if there are any constraints on $a^\mu$ or $\Lambda^{\mu}_{\;\nu}$. The Jacobian for the general transformation is given by
\begin{equation*}
\begin{aligned}
    \frac{\partial\tilde{x}^\rho}{\partial x^\mu}&=\frac{\partial}{\partial x^\mu}(a^\rho+\Lambda^{\rho}_{\;\nu} x^\nu) \\
    &=\Lambda^\rho_{\;\nu}\partial_\mu x^\nu \\
    &=\Lambda^\rho_{\;\nu}\eta_\mu^{\;\nu} \\
    &=\Lambda^\rho_{\;\nu}
\end{aligned}
\end{equation*}
If we then plug this into equation \ref{poincareconstraint}, we are left with 
\begin{equation} \label{lorentz}
\eta_{\mu\nu}=\Lambda^\rho_{\;\mu}\Lambda^\sigma_{\,\nu}\eta_{\rho\sigma}
\end{equation}
You'll notice that the constant vector $a^\mu$ dropped out of our equation. This means we are free to choose it to be anything we like. These will correspond to translations in spacetime and we learn that we can translate by any amount we like.
The solutions of \ref{lorentz} form the Orthogonal Group $O(D-1,1)$, which was discussed in section \ref{group}. They form a subgroup of the Poincar\'e Group, which is also known as the \textit{Lorentz Group}. These transformations correspond to spatial rotations and boosts, which can be thought of as a type of rotation that rotate between space and time. 

We can go further and consider subgroups of the Lorentz Group. To do this, take the determinant of both sides of equation \ref{lorentz}
\begin{equation*}
\begin{aligned}
    \det(\eta)&=\det(\Lambda^T\eta\Lambda) \\
    \det(\eta)&=\det(\Lambda^T)\det(\eta)\det(\Lambda) \\
    1&=\det(\Lambda^T)\det(\Lambda) \\
    1&=\det(\Lambda)^2
\end{aligned}
\end{equation*}
We find that for all orthogonal matrices, $\Lambda$, we have $det(\Lambda)=\pm 1$. Singling out only the subset of orthogonal matrices with positive determinant allows us to create further subgroups.

\vspace{0.5cm}

\noindent \underline{Special Orthogonal Group, $SO(1,D-1)$}: $\det(\Lambda)=1$
This subgroup contains only the transformations that have determinant $+1$. It turns out that this group is all of the transformations which preserve orientation. These transformations are said to be \textit{Proper}.

\vspace{0.5cm}

\noindent \underline{Orthochronous Orthogonal Group, $O^+(1,D-1)$}: $\Lambda^0_0=+1$
This is the group of Lorentz transformations that preserve the direction of time. 

\vspace{0.5cm}

In physics, we primarily concern ourselves with the \textit{Proper Orthochronous Lorentz Group}, $SO^+(1,D-1)$. When a physicist speaks of the Lorentz group, it is often this subgroup they are referring to. Note, for instance, that it is a property of all QFTs that they must have symmetry under this subgroup rather than the full Orthogonal Group.

However, we sometimes also want to ask if our theory is invariant under the not proper or not orthochronous transformations. What are these transformations?

\vspace{0.5cm}

\noindent \underline{Parity}: Flips the orientation of space.

\begin{equation*}
P=
\begin{bmatrix}
    1 & 0 & 0 & 0 \\
    0 & -1 & 0 & 0 \\
    0 & 0 & -1 & 0 \\
    0 & 0 & 0 & -1
\end{bmatrix}
\end{equation*}

\noindent \underline{Time Reversal}: Changes the direction of time.

\begin{equation*}
T=
\begin{bmatrix}
    -1 & 0 & 0 & 0 \\
    0 & 1 & 0 & 0 \\
    0 & 0 & 1 & 0 \\
    0 & 0 & 0 & 1
\end{bmatrix}
\end{equation*}
We can also apply both of these in a single transformation.

\begin{equation*}
PT=
\begin{bmatrix}
    -1 & 0 & 0 & 0 \\
    0 & -1 & 0 & 0 \\
    0 & 0 & -1 & 0 \\
    0 & 0 & 0 & -1
\end{bmatrix}
\end{equation*}

An important property of these last three transformations that distinguishes them from the proper orthochronous transformations is that they are not connected to the identity. Flipping spacial orientation and time reversal are binary operations, you either do them or you do not, you cannot do them by some small amount. This means you cannot start with the identity transformation and consider an infinitesimal spacial orientation flip or time reversal. This leads to some terminology that you should be familiar with. When people talk about the \textit{connected part of the Lorentz group}, they are referring to the \textit{set of Lorentz transformations that are connected to the identity transformation}, i.e. the proper orthochronous Lorentz transformations.

\subsubsection{The Poincare Algebra} \label{poincarealg}

We would now like to find the generators of the proper orthochronous transformations because from those we can derive what is known as the \textit{Poincar\'e Algebra}. This requires considering infinitesimal transformations, as was discussed briefly in section \ref{lie}. Consider $$x^\mu\rightarrow x^\mu+\xi^\mu$$
where $\xi^\mu$ is some infinitesimal transformation. We will want to substitute this into equation \ref{poincareconstraint}, but to do so we first must find $\frac{\partial\tilde{x}^\mu}{\partial x^\rho}$
\begin{equation*}
\begin{aligned}
    \frac{\partial\tilde{x}^\mu}{\partial x^\rho}&=\frac{\partial}{\partial x^\rho}(x^\mu+\xi^\mu(x)) \\
    &=\eta^\mu_\rho+\partial_\rho\xi^\mu
\end{aligned}
\end{equation*}
Now, substituting into equation \ref{poincareconstraint} gives us $$\eta_{\mu\nu}=(\eta^\rho_\mu+\partial_\mu\xi^\rho)(\eta^\sigma_\nu+\partial_\nu\xi^\sigma)\eta_{\rho\sigma}$$
If we expand this and only keep terms that are linear in the derivative of $\xi^\mu$, then we get $$\eta_{\mu\nu}=\eta_{\mu\nu}+\partial_\nu\xi_\mu+\partial_\mu\xi_\nu$$
From this we can conclude that 
\begin{equation} \label{killing}
\partial_\nu\xi_\mu+\partial_\mu\xi_\nu=0
\end{equation}
This is known as the \textit{Killing equation}. The solutions to this equation will give us our infinitesimal Poincar\'e transformations. We would like to massage this equation a bit to get some information from it. What we will do is consider taking the derivative, $\partial_\rho$, and permuting the indices to come up with multiple equations, all equal to $0$. Since they are all equal to $0$ we can then add them all together. Following through with this process looks like this
\begin{equation*}
\begin{aligned}
    &\quad \partial_\rho\partial_\mu\xi^\nu+\partial_\rho\partial_\nu\xi_\mu \\
    &+(\partial_\mu\partial_\nu\xi^\rho+\partial_\nu\partial_\mu\xi_\rho) \\
    &-(\partial_\nu\partial_\rho\xi^\mu+\partial_\mu\partial_\rho\xi_\nu)=0
    \end{aligned}
\end{equation*}
Using the fact that partial derivatives commute, many of these terms will cancel, leaving us with 
\begin{equation}
\partial_\mu\partial_\nu\xi_\rho=0
\end{equation}
This tells us that the second derivative of $\xi$ vanishes, which means that $\xi$ must be at most linear in $x$. From this we can write the general solution for $\xi$ $$\xi_\rho=a_\rho+\omega^{\;\sigma}_\rho x_\sigma$$
We then want to plug this result into equation \ref{killing} to see if there are any constraints on $a_\rho$ or $\omega^{\;\sigma}_\rho$.
\begin{equation*}
\begin{aligned}
    \partial_\nu(a_\mu+\omega_\mu^{\;\sigma} x_\sigma)+\partial_\mu(a_\nu+\omega_\nu^{\;\rho} x_\rho)&=0 \\
    \omega_\mu^{\;\sigma}\eta_{\nu\sigma}+\omega_\nu^{\;\rho}\eta_{\mu\rho}&=0 \\
    \omega_{\mu\nu}+\omega_{\nu\mu}&=0 \\
    \omega_{\mu\nu}&=-\omega_{\nu\mu}
\end{aligned}
\end{equation*}
We find that $\omega$ must be anti-symmetric. Therefore, the most general solution is $$\xi_\rho=a_\rho+\omega_\rho^{\;\sigma} x_\sigma$$ where $a_\rho$ represents translations and $\omega_\rho^{\;\sigma}$ represents rotations and boosts and is an anti-symmetric tensor.

One might wonder how many independent parameters there are in this equation. Working in 3+1 spacetime dimensions, it is straightforward to see that $a^\mu$ has 4 independent parameters, one for each of the 3 spatial translations and one for time translation. Without any constraints, $\omega^{\;\rho}_\sigma$ would have 16 independent parameters. However, we know that it must be anti-symmetric, which means it only has 6 independent parameters, 3 for each of the spatial rotations, and 3 for each of the boosts.

These are the generators of the Lorentz Group. We can represent these generators as differential operators acting on our spacetime. This is hopefully familiar from your Quantum Mechanics course, where you learned that momentum is the generator of translations and is represented by a derivative operator. You can convince yourself that this makes sense by exponentiating your generator, which you expect should lead to a finite translation and then acting on a function with it. Let's see explicitly how this works for the case of translations.

We recall that the generator of translations is momentum and it can be expressed as $$\hat{P}_\mu=-i\partial\mu$$ where the factor of $-i$ is conventional in order to guarantee that the operator is Hermitian. If we would like to produce a finite translation, then we exponentiate this infinitesimal generator of translations $$e^{-ia^\mu\hat{P}_\mu} f(x)=e^{-ia^\mu(-i\partial_\mu)}f(x)=e^{-a^\mu\partial_\mu}f(x)$$
To see that this does what we want, we can Taylor expand
\begin{equation*}
\begin{aligned}
    e^{-a^\mu\partial_\mu}f(x)&=[1-a^\nu\partial_\nu+\frac{1}{2}(-a^\rho\partial_\rho)^2+...]f(x) \\
    &=f(x)-a^\nu\partial_\nu f(x)+\frac{1}{2}(a^\rho\partial_\rho)^2f(x)+\dots \\
    &=f(x-a)
\end{aligned}
\end{equation*}
where we used the fact that the second line is just the power series representation of the third line.

Similarly, the generators of rotations can be found to be $$\hat M_{\mu\nu}=i(x_\mu\partial_\nu-x_\nu\partial_\mu)$$ Again, the factor of $i$ is conventional.

With these representations of our generators in hand, we can compute the Poincar\'e Algebra, which is nothing more than the commutation relations between all of the generators.
\begin{equation*}
\begin{aligned}
    [\hat P_\mu,\hat P_\nu]&=(-i\partial_\mu)(-i\partial_\mu)-(-i\partial_\nu)(-i\partial_\mu) =0 \\
    [\hat P_\mu,\hat M_{\nu\rho}]&=(-i\partial_\mu)i(x_\nu\partial_\rho-x_\rho\partial_\nu)-i(x_\nu\partial_\rho-x_\rho\partial_\nu)(-i\partial_\mu) \\
    &=+i(\eta_{\mu\nu}\hat P_\rho-\eta_{\mu\rho}\hat P_\nu) \\
    [\hat M_{\mu\nu},\hat M_{\rho\sigma}]&=i(x_\mu\partial_\nu-x_\nu\partial_\mu)i(x_\rho\partial_\sigma-x_\sigma\partial_\rho)-i(x_\rho\partial_\sigma-x_\sigma\partial_\rho)i(x_\mu\partial_\nu-x_\nu\partial_\mu) \\
    &=i(\eta_{\nu\rho}\hat M_{\mu\sigma}-\eta_{\nu\sigma}\hat M_{\mu\rho}-\eta_{\mu\rho}M_{\nu\sigma}+\eta_{\mu\sigma}M_{\nu\rho}) 
\end{aligned}
\end{equation*}
where we used the fact that partial derivatives commute and $\partial_\mu x_\nu =\eta_{\mu\nu}$.

\subsection{Quantum Field Theory} \label{QFT}
In this primer, we will mostly be focused on Quantum CFTs. In fact, we will generally just say CFT and it is implied that we mean a quantum theory. In this case, a CFT is really just a special type of Quantum Field Theory (QFT)\footnote{There are many standard introductions to QFT, including \cite{srednicki,peskin,weinberg}}. Specifically, it is a QFT with conformal symmetry. As the target audience of this primer is undergraduate and early-career graduate students, we expect most readers will have little to no experience with QFT. Happily, the power of symmetry will allow us to make a lot of progress in CFT without all the technical expertise that comes out of a full class in QFT. In this section, we will highlight some important aspects of QFT that will be especially important to future discussions.

Before we can talk about QFT, we need to get on the same page about Classical Field Theory. When a Physicist talks about a \textit{field}, all they mean is a mathematical object that takes a position as its input. The output can be a variety of different types of objects. For instance a scalar field takes a position and outputs a scalar, while a vector field takes a position and outputs a vector. Fields are likely more familiar than you might think. Let's see a couple examples.

\begin{itemize}
    \item Temperature is an example of a scalar field. The temperature of a room can be described by $T(t,\vec{x})$. So, given the position $\vec{x}$ at time $t$, the function $T$ returns the temperature at that point, which is a scalar value.
    \item The electric field is an example of a vector field. The electric field can be described by $\vec{E}(t,\vec{x})$. So, given the position $\vec{x}$ at time $t$, the function $\vec{E}$ returns the electric field at the point. The electric field carries both magnitude and direction and is therefore a vector. 
\end{itemize}

So, what is QFT? It is the marriage of Quantum Mechanics and Special Relativity. It turns out that in order for these two fundamental theories to be consistent, we need this new framework. As an example of the need for this, recall your QM course. There, position was an operator, but time was simply a parameter. In SR, we learn that we should treat space and time on equal footing. So, even at this very basic level, we see that there are problems. These are solved by QFT.

If we want to quantize a classical field theory, our fields will become operators. These operators are similar to the operators from your QM course, except they are now \textit{local operators} because they depend on position. In QFT, we will have different types of operators and there are multiple ways these operators can be classified. For instance, operators can be classified based on the way they transform under Lorentz transformations. We can have scalars, vectors, spinors, etc. A second operator classification that will be very important in CFT is \textit{Operator Dimension}.

\subsubsection{Operator Dimension} \label{dimension}
First, note that this is very different from spacetime dimension. When we use the word dimension here, we mean dimension like in dimensional analysis. It is very important to keep track of what type of dimension we are talking about (spacetime dimension vs operator dimension). We will use standard convention and use $D$ for spacetime dimension and $\Delta$ for operator dimension. To make it somewhat more confusing, in QFT, there are several terms that can be used somewhat interchangeably to describe operator dimension, for instance scaling dimension, mass dimension, or engineering dimension.

The operator dimension will tell us how our operators transform under spacetime dilatations, which are a re-scaling of our coordinates $x^\mu\rightarrow\tilde x^\mu=\lambda x^\mu$, where $\lambda$ is a constant known as the scale factor. This is where the name scaling dimension comes from.

How do we know the operator dimension? This is where the term mass dimension comes in handy. For familiar variables (e.g. space, time, momentum, energy) we can determine their operator dimension by comparing them with mass. To see what we mean by this, start by setting $\hbar=c=1$, as is generally done in QFT (after all, those are just constants and making this choice is equivalent to working with a particular set of units). With this in mind, we can review some familiar equations and see how the variables on the different sides of the equation are related dimensionally (using dimensional analysis). We will call the dimension of mass one, $[m]=+1$ and count the dimension in terms of powers of mass.

\begin{itemize}
        \item We know that $E=mc^2=m$. This implies $[E]=[m]=+1$. This result should make sense as we know from the theory or relativity that mass is just a particular form of energy.
        \item You can also determine what the mass dimension of length is. One way to do this is by considering the energy of a photon which is given by $E=\frac{h}{\lambda}$. From there we can conclude that the dimensional relationship is $[E]=[\frac{1}{x}]$ and therefore $[x]=-1$. The intuition for this is that as the wavelength decreases, the energy increases so there is an inverse relationship between energy and length, hence the dimension of $-1$.
        \item Since we are considering relativistic systems, space and time are on the same footing, which means time must have the same mass dimension as space (length), therefore $[t]=-1$.
        \item As a last example, let us consider the mass dimension of momentum. Thinking about the relativistic energy equation, $E^2=(mc^2)^2+(pc)^2$, we know the units on the left must match the units on the right, which means all 3 terms must have the same units. In the same way, the dimension of each side must mach, so $[E]=[m]=+1 \implies [p]=+1$.
\end{itemize}
All operators in a QFT have an associated dimension. If you know the Lagrangian for the system you are considering, you can determine the dimension of the field. Let's see how to do this. Imagine you are given the action
$$S=\int d^Dx\mathcal{L}$$
As previously discussed, the dimension of position is given by $[x]=-1$, which means $d^D x$ has dimension $[d^D] x=-D$. Our action must be dimensionless. Therefore, the Lagrangian must have dimension $[\mathcal{L}]=+D$. This means that every term in our Lagrangian will have dimension $+D$. 

Let's do an example. The Lagrangian will always have some kinetic energy term. For a scalar field theory, this term is given by $$\mathcal{L}_{\mathrm{kin}}= \partial_\mu\phi\partial^\mu\phi$$
Therefore, we know $$[\partial_\mu\phi\partial^\mu\phi]=+D$$
Derivatives have dimension of inverse length, therefore $[\partial_\mu]=+1$. Putting this all together, we are left with $$[\phi]=\frac{D-2}{2}$$
Given a Lagrangian, one can always compute the mass dimension of the operators in this way. Therefore, if we consider making a dilatation transformation (re-scaling) of your system, we can see how the operator transforms. This will be critical in CFT, where our systems will have a symmetry under this transformation.

\subsubsection{Renormalization Group (RG) Flows} \label{rgflow}
In classical physics, it is common to change the way you describe how your system behaves depending on the energy scale of the system. A good example of this is water. The behavior of water at room temperature and sea level pressure is accurately described by fluid mechanics, where the fact that water is made up of discrete $H_2O$ molecules is ignored. However, if you probe at higher energy scales (for instance when it turns into a vapor), you can see that the water is made up of discrete molecules and the continuous model of fluid mechanics breaks down.

Another example we could consider is the classic bead-on-a-wire problem that lends itself nicely to Lagrangian mechanics. When this problem is visited in a mechanics course, the bead is constrained to move along a path that follows the wire. This constraint brings with it an implicit restriction on the energy levels of the bead. In reality, if the bead has high enough energy it will deform the wire as opposed to moving along its path, and at higher energy still, the bead would break the wire altogether. 

In quantum theories, one must consider all possible outcomes. This is explicit in the \textit{path integral approach}, where you sum over all possible trajectories. It seems then that if we consider quantum versions of these systems, then we might need to account for all of the different types of behaviors at different energy scales. Along these lines, when writing the action for a system, you begin by assuming that all possible interactions are present. Whether an interaction is possible or not is determined by whether or not symmetry permits the interaction (for instance, all terms in a QFT must be Lorentz Invariant). Even with symmetry considerations, including all possible interactions leads to an action which has infinitely many terms, which seems problematic. For a scalar field, for instance, some terms we would need to consider are
\begin{equation*}
\begin{aligned}
    &\phi^2(x),\phi^3(x),...,\phi^n(x),... \\
    &\partial_\mu\phi\partial^\mu\phi,(\partial_\mu\phi\partial^\mu\phi)^2... \\
    &\phi\partial^2\phi,...
\end{aligned}
\end{equation*}

Happily, it will turn out that not all terms will be important at finite energies. This leads to an additional way of classifying operators
\begin{itemize}
    \item Relevant Operator: $\Delta<D$
    \item Irrelevant Operator: $\Delta>D$
    \item Marginal Operator: $\Delta=D$
\end{itemize}
where $\Delta$ is the operator dimension (note that here we mean the dimension of the full term in the Lagrangian, not just the dimension of the scalar field itself) and $D$ is the spacetime dimension.

We can think about this conceptually. The terms in the Lagrangian tell us what types of interactions are allowed. Some interactions take a lot of energy to excite (just like how it would take a lot of energy for the bead to break the wire and fly off in the classical example). Therefore, although they are allowed Quantum Mechanically, the probability just becomes too small to worry about. We can see this mathematically by considering the \textit{Beta Function}. Each term in the Lagrangian comes with a coupling constant $g_i$. This constant tells you how strong the interaction is. The Beta Function will tell you how this coupling changes with energy scale $\mu$. $$\beta_i=\frac{\partial g_i}{\partial \log \mu}$$

For Relevant Operators, the Beta Function dictates that the strength of the interaction grows at low energies. On the other hand, the strength of Irrelevant Operators decreases, which is why they are said to be irrelevant. Ultimately, you can write an \textit{Effective Action} that only includes the relevant and marginal operators, which leaves only a finite number of terms.
 
Analyzing how the description of your system changes with energy is known as the \textit{Renormalization Group (RG) Flow}\footnote{Please note that the Renormalization Group is not a group in the sense of section \ref{group}. This is an unfortunate nomenclature.}. Analyzing this flow allows us to consider the full space of allowed Quantum Field Theories. We will discuss the fact that CFTs live at special places along the RG flow in section \ref{intro}.

\subsubsection{Correlation Functions} \label{correlationfunct}
Our final topic in this QFT review is on Correlation Functions. These are similar to Expectation Values, which you should be familiar with from your Quantum Mechanics course. These are objects of the form $$\langle 0|\phi_1(x_1)\phi_2(x_2)\dots \phi_n(x_n)|0\rangle=\langle \phi_1(x_1)\phi_2(x_2)\dots \phi_n(x_n)\rangle$$ where $|0\rangle$ is the ground state of your system and the $\phi_i(x_i)$ are local operators. In general, the operators may have any type of Lorentz representation (e.g. scalar, vector, etc) or operator dimension.

They are called correlation functions because they can tell you about how different points are correlated with regard to a certain operator. These are also referred to as \textit{$n$-point} functions. Where the $n$ refers to the number of operators. For instance, in particle physics, a three-point correlator might be related to the amplitude for a particle to decay (starting at one point in spacetime and then decaying into particles at two other points). A four-point function might represent a scattering amplitude. As a second example, you could consider the Ising Model, which is made up of spins on a lattice. There, the correlation function might be telling us about how much the spins at different points are correlated.

Correlation Functions are very important in any Quantum Theory, as they are used to make predictions that can be experimentally verified. In CFT, we will see that one-, two-, and three-point functions can be completely determined up to a set of constants by enforcing symmetry alone, which really tells us about the power of symmetry!

\section{What is Conformal Field Theory and Why Should You Care?} \label{intro}
A Conformal Field Theory (CFT) is simply a field theory with conformal symmetries. CFTs can be classical or quantum. Interestingly, it is possible to have a classical field theory with conformal symmetries that has these symmetries broken when the theory is quantized. The breaking of these symmetries are known as \textit{anomalies}.

In section \ref{confsymm}, the Conformal Group is described and you will see that it is larger than the Poincar\'e Group, which was discussed in section \ref{poincare}. In fact, the Poincar\'e Group is a subgroup of the Conformal Group. This large amount of symmetry gives us a lot of power in computing physical quantities, like expectation values, for CFTs. In fact, we will see in section \ref{correlators} that many physical quantities are fully determined (up to a set of constants that are associated with your specific system) by conformal symmetry. Furthermore, we can compute these things without needing much of the mathematical framework that comes out of more general Quantum Field Theory (QFT). In this way, CFT is more accessible than QFT.

That's all great, but why should you actually care about CFT? We will give four answers to this question and there are certainly more.

\hspace{1cm}

\noindent 1. \textit{To better understand the space of all Quantum Field Theories.}

If you are examining a particular physical system, you would do so using a particular QFT that is used to describe that system. For instance, Quantum Electrodynamics is a QFT that describes the quantum theory of Electromagnetism and has a particular associated Lagrangian. QFT provides the framework for describing a large variety of very different physical systems. Thinking about it this way, one might wonder about all of the different particular QFTs that are out there. Can you have any types of fields you want with any types of interactions? When we say the space of all QFTs, we mean the space of all possible fields and interactions that are allowed by, e.g., symmetry or unitarity.

One of the ways that people organize thinking about the space of all QFTs is through the coupling parameters, $g_i$, which are a measure of how strongly fields are going to interact. Theories will generally have multiple coupling parameters and one can think about what will happen if these various coupling parameters change. In QFT, an interesting thing happens and that is that the values of these coupling parameters can change depending on the energy level at which you are probing your system. This is quantified using the beta function, which was introduced in \ref{rgflow}. In QFT, for every coupling parameter, $g_i$ there is an associated beta function $\beta_i$ such that $$\beta_i=\frac{\partial g_i}{\partial \log \mu}$$ where $\mu$ is the energy scale. We see the beta function quatifies the rate of change of the coupleings as you change the energy. What happens when this rate goes to zero? Then, the couplings no longer depend on scale. This is where we find scale invariant theories. We will learn that CFTs are scale-invariant, so this is one way to find CFTs.  What this tells us is that CFTs live within the space of all QFTs. Specifically, they exist at the critical points of the beta function.

With this all in mind, one can imagine starting with a CFT. We will see that the power of symmetry allows us to make a lot of progress in analyzing these. One can then perturb the CFT away from the critical point. In this way, one can use perturbative methods to solve for physical quantities in non-conformal QFTs, which are more challenging to analyze directly, by starting with a CFT.

\hspace{1cm}

\noindent 2. \textit{To understand critical phenomena (phase transitions) in statistical systems}

There are many physical systems in condensed matter physics and statistical mechanics that are described by a CFT. Generally these systems become conformal at their critical points.

A good specific example of this is the Ising Model, which describes ferromagnetism. This model has a critical point where it becomes scale invariant and, in that case, is described by a CFT. Interestingly, the phase transition of water is actually also described by a 3D Ising model and therefore is described by a CFT.

\hspace{1cm}

\noindent 3. \textit{String Theory is a 2 dimensional CFT}

In string theory\footnote{See, e.g. \cite{polchinski}}, it is postulated that at the most fundamental level, particles do not exist as points, but rather as strings. A string is just a 1D line and as the string propagates through time, it creates a 2D sheet, which is called the string's worldsheet. The QFT that lives on the worldsheet is conformal. Therefore, CFT is the underlying framework for string theory. At this time, many believe that string theory is the best candidate for Quantum Gravity, which aims to unite Quantum Mechanics and General Relativity.

\hspace{1cm}

\noindent 4. \textit{AdS/CFT Duality (Quantum Gravity)}

The AdS/CFT duality \cite{Maldacena_1999,Gubser_1998,witten1998anti} is an incredible tool for helping us to understand Quantum Gravity. Here, AdS is the description of Quantum Gravity in Anti-de Sitter space, where Anti-de Sitter space refers to one of the maximally symmetric solutions to Einstein's equations. As it turns out, seemingly magically, this AdS space is dual to a CFT without gravity. This means that every observable (e.g. scattering amplitude) in Anti-de Sitter space can mathematically be mapped to an observable (e.g. a correlation function) in a CFT. Therefore, you can compute various quantities in CFT and use them to learn about Quantum Gravity. There are many excellent reviews of this, including \cite{Penedones_2016,Aharony_2000,McGreevy_2010,dhoker2002supersymmetric,polchinski2010introduction}.

\section{Conformal Symmetry} \label{confsymm}

\subsection{The Conformal Group}

Conformal Transformations are the set of transformations that leave the metric $\eta_{\mu\nu}$ invariant up to an overall scale factor. For a transformation $x^\mu\rightarrow\tilde{x}^\mu(x)$, the metric transforms as follows:
    \begin{equation*}
\Tilde{\eta}_{\mu \nu}=\frac{\partial \Tilde{x}^\rho}{\partial x^\mu}\frac{\partial \Tilde{x}^\sigma}{\partial x^\nu}\eta_{\rho \sigma} 
    \end{equation*}
A transformation is said to be conformal if 
    \begin{equation} \label{conf}
            \Tilde{\eta}_{\mu \nu}=\Omega^2(x)\eta_{\mu \nu}
    \end{equation}
where $\Omega^2(x)$ is the position-dependant scale factor (note that it is standard convention to define the scale factor as $\Omega^2(x)$, rather than just $\Omega$(x)). We see that these are the transformations that re-scale your spacetime coordinates by an amount that can vary by location. However, not any re-scaling is allowed. Conformal Transformations are those that preserve angles. The name is the same as is used in cartography, where a conformal map is a projection of the Earth onto a flat surface that preserves angles. This is useful for sailors at sea who care more about going in the right direction than knowing how much further they have to go. 

The set of transformations that do this are:
    \begin{itemize}
        \item \textbf{Lorentz Transformations}
        $$\Tilde{x}^\mu=\Lambda^\mu{\,_\nu} x^\nu$$ where $\Lambda^\mu{\,_\nu}$ is a constant matrix that satisfies $\eta_{\mu\nu}=\Lambda^\rho_{\,\mu}\Lambda^\sigma_{\,\nu}\eta_{\rho\sigma}$. These transformations can be spacial rotations, boosts to frames moving at different velocities, or some combination of these.
        \item \textbf{Translations} $$\Tilde{x}^\mu=x^\mu+a^\mu$$ where $a^\mu$ is a constant vector that translates all points together in spacetime. Notice that the Lorentz transformations plus translations form the Poincar\'e group. These are the subset of the conformal group where $\Omega(x)=1$. The Poincar\'e group was discussed in detail in section \ref{poincare}.
        \item \textbf{Scale Transformations (aka Dilatations)} $$\Tilde{x}^\mu=\lambda x^\mu$$ where $\lambda$ is a constant. Dilatations re-scale your spacetime by the same amount everywhere. It is the special case where your scale factor $\Omega(x)$ is a constant.
        \item \textbf{Inversion} $$\Tilde{x}^\mu=\frac{x^\mu}{x^2}$$ This takes points that are close to the origin and sends them far away, and vice versa. Notice this is different from the previous transformations in that there is just one way to invert (i.e. there is no constant vector or matrix that you get to choose, there is just one inversion).
        \item \textbf{Special Conformal Transformations (SCT)} $$\Tilde{x}^\mu=\frac{x^\mu+b^\mu x^2}{1+b^2x^2+2b_\rho x^\rho}$$ where $b^\mu$ is a constant vector. These transformations are the most challenging to visualize. They can be thought of as an inversion, follwed by a translation, followed by another inversion.
        \begin{center}
        SCT $\iff$ Inversion $\rightarrow$ Translation $\rightarrow$ Inversion
        \end{center}
        We can see this explicitly. Start with an inversion, $$\Tilde{x}^\mu=\frac{x^\mu}{x^2}$$

        Followed by applying a translation, $$\Tilde{\Tilde{x}}^\mu=\Tilde{x}^\mu+b^\mu=\frac{x^\mu}{x^2}+b^\mu$$

        Followed by a final inversion.
        \begin{equation*}
        \begin{aligned}
        \Tilde{\Tilde{\Tilde{x}}}^\mu=\frac{\Tilde{\Tilde{x}}^\mu}{\Tilde{\Tilde{x}}^2}&=\frac{\frac{x^\mu}{x^2}+b^\mu}{(\frac{x^\nu}{x^2}+b^\nu)(\frac{x_\nu}{x^2}+b_\nu)} \\
        &=\frac{x^\mu+b^\mu x^2}{1+b^2x^2+2b_\rho x^\rho}
        \end{aligned}
        \end{equation*}
        As promised, we see that applying this series of transformations is the same as applying the Special Conformal Transformation.
        \item \textbf{Parity} $$\Tilde{x}^\mu=-x^\mu,\,\,\,\mu=1,2,3,\dots$$ this flips your spacial points across their axis, like a mirror reflection.
        \item \textbf{Time-Reversal} $$\Tilde{x}^0=-x^0$$ This sends positive time to negative time and vice versa. It is similar to parity, but for time time coordinate.
    \end{itemize}

All seven of the aforementioned transformations are conformal, in that they satisfy eqn. \ref{conf}. However, we do not always want to include all of these. Just like was the case with Poincar\'e transformations, as discussed in section \ref{poincare}, we oftentimes only want to consider the set of transformations that can be taken to be infinitesimally close to the identity. Parity, Time-Reversal, and Inversion are all transformations that are fixed, i.e. they do not have any parameters that you get to choose. Therefore, there is no way to consider a small transformation for these types. This set of transformations form a group (group theory is reviewed in section \ref{group}). If we only want to consider the first four transformations (Lorentz, translations, dilatations, and SCTs), then we call this the connected part of the group, as these are the transformations that can be connected to the identity.

\subsection{The Conformal Algebra}

In the previous discussion, we simply stated what the conformal transformations are without any proof. In this section, we will find the transformations in the connected part of the conformal group by considering infinitesimal transformations. In this way, we can find the generators of the transformations and ultimately derive the Conformal Algebra. Again, we want the set of transformations that preserve the metric up to a local change in scale.
  \begin{equation*} \tag{\ref{conf}}
\Tilde{\eta}_{\mu \nu}=\frac{\partial \Tilde{x}^\rho}{\partial x^\mu}\frac{\partial \Tilde{x}^\sigma}{\partial x^\nu}\eta_{\rho \sigma} =\Omega^2(x)\eta_{\mu \nu}
    \end{equation*}

We will start by considering an infinitesimal transformation. Any infinitesimal transformation can be expressed as follows: $$x^\mu\rightarrow\tilde{x}^\mu=x^\mu+\xi^\mu$$ where $\xi^\mu$ is taken to be the infinitesimal parameter. This transformation has an associated Jacobian matrix given by $$\frac{\partial\tilde{x}^\rho}{\partial x^\mu}=\eta_\mu^\rho+\partial_\mu\xi^\rho$$
Since the transformation is infinitesimal, it will correspond with a scale factor $\Omega(x)$ that differs from one infinitesimally. We can therefore also expand our scale factor as $\Omega(x)=1+\omega(x)+...$, where $\omega(x)$ is small. Substituting these into this into equation \ref{conf}, we have
\begin{equation*}
\begin{aligned}
    \eta_{\rho\sigma}(\eta_\mu^\rho+\partial_\mu\xi^\rho)(\eta_\nu^\sigma+\partial_\nu\xi^\sigma)&=(1+\omega(x)+...)^2\eta_{\mu\nu} \\   =\eta_{\mu\nu}+\partial_\mu\xi_\nu+\partial_\nu\xi_\mu+O((\partial_\mu\xi_\sigma)^2)&=\eta_{\mu\nu}+2\omega(x)\eta_{\mu\nu}+O(\omega(x)^2)
\end{aligned}
\end{equation*}
Terms of order $O((\partial_\mu\xi^{\sigma})^2)$\footnote{Some readers may be unfamiliar with this notation. Here the O is representing Order. As an explicit example for how this is used, if we take $x$ to be small, anything $x^2$ is really small, and anything $x^3$ is even smaller. So, we can do an approximation where we drop terms $O(x^2)$, which means we drop any terms that have powers of $x$ greater than or equal to 2.} and $O(\omega(x)^2)$ or higher are very small and therefore can be dropped. We are left with
\begin{equation} \label{infconf} \partial_\mu\xi_\nu+\partial_\nu\xi_\mu=2\omega(x)\eta_{\mu\nu}
\end{equation}
This alone is not very enlightening. To see what more we can learn, we will take the trace of both sides, i.e. we will contract the indices on both sides with $\eta^{\mu\nu}$

\begin{equation*}
\begin{aligned}
    \eta^{\mu\nu}(\partial_\mu\xi_\nu+\partial_\nu\xi_\mu)&=2\omega(x)\eta_{\mu\nu}\eta^{\mu\nu} \\
    2\partial^\mu\xi_\mu&=2D\omega(x) 
\end{aligned}
\end{equation*}

\begin{equation} \label{infscale}
    \omega(x)=\frac{1}{D}\partial^\mu\xi_\mu(x)
\end{equation}

If we now plug \ref{infscale} into \ref{infconf}, we are left with a differential equation only involving only $\xi_\mu$. 
\begin{equation} \label{confkilling}   \partial_\mu\xi_\nu+\partial_\nu\xi_\mu=\frac{2}{D}\partial^\sigma\xi_\sigma\eta_{\mu\nu}
\end{equation}
This is known as the \textit{conformal Killing equation}. Its solutions will be the generators of the Conformal Group. By taking two derivatives of both sides of this equation, considering permutations of the indices, and combining, one can show that \ref{confkilling} leads to $$\partial_\mu\partial_\nu\partial_\lambda\xi_\rho=0$$ This makes it clear that the most general solution is a polynomial in $x$ of order two\footnote{This is like how if you have a single variable scalar function, $f(x)$, and you know $d^3f/dx^3=0$, then the solution is a polynomial in $x$ with powers no higher than 2, e.g. $Ax^2+Bx+C$}. The most general solution to this is therefore given by $$\xi_\mu=a_\mu+\omega_\mu^{\,\nu} x_\nu+\lambda x_\mu+b_\mu x^2-2x^\mu b_\rho x^\rho$$ Here, $a_\mu$, $b_\mu$, and $\lambda$ are arbitrary constants and $\omega_{\mu\nu}=-\omega_{\mu\nu}$. The signs and factors of two are conventional for later convenience. In 4 spacetime dimensions, there are 15 free parameters, which represent the following transformations:

\begin{itemize}
    \item $a_\mu \rightarrow$ 4 spacetime translations
    \item $b_\mu \rightarrow$ 4 special conformal transformations
    \item $\lambda \rightarrow$ 1 dilatation
    \item $\omega_{\mu\nu} \rightarrow$ 3 rotations and 3 boosts
\end{itemize}
More generally, in $D$ spacetime dimensions, there are $(D+2)(D+1)/2$ free parameters. 

These transformations can be represented as differential operators acting on your spacetime. This might be familiar from Quantum Mechanics, where one sees that momentum, which is the generator of translations, can be represented by a derivative. All of the other generators of conformal transformations can similarly have representations as differential operators. 

\begin{itemize}
    \item Momentum: $\hat{P}_\mu=-i\partial_\mu$
    \item Rotations and Boosts: $\hat{M}_{\mu\nu}=i(x_\mu\partial_\nu-x_\nu\partial_\mu)$
    \item Dilatations: $\hat{D}=ix^\mu\partial_\mu$
    \item SCT: $\hat{K}_\mu=i(x^2\partial_\mu-2x_\mu x^\nu\partial_\nu)$
\end{itemize}
where the included factors of $i$ are the standard convention. It is important to note that this is not the only convention that you will come across and one must always be sure to make note of what convention is being used. As this is the standard convention, it is the one we will use in most of this paper. However, we will consider an alternative convention in section \ref{Primaries}, as it will make a helpful analogy explicit.

Remember that these generators lead to infinitesimal transformations. If you want a finite transformation, you simply need to exponentiate, which again should be familiar from Quantum Mechanics where you exponentiate momentum to translate in space and you exponentiate your Hamiltonian to translate in time.

With this representation of the generators in hand, we can derive the Conformal Algebra. We do this by computing commutators of different combinations of the generators. This is exactly as we did in section \ref{poincare} for the Poincar\'e Algebra. We will not do this for all generators, but the reader is encouraged to work through the computations on their own. As an example, we compute the commutator of Momentum and Dilatations.

\begin{equation*}
\begin{aligned}
    [\hat{P}_\mu,\hat{D}]&=(-i\partial_\mu)(-ix^\nu\partial_\nu)-(-ix^\nu\partial_\nu)(-i\partial_\mu) \\
    &=\partial_\mu(x^\nu\partial_\nu)+(x^\nu\partial_\nu)\partial_\mu \\
    &=\eta_\mu^{\nu}\partial_\nu+x^\nu\partial_\mu\partial_\nu-x^\nu\partial_\nu\partial_\mu \\
    &=\eta_\mu^{\nu}\partial_\nu \\
    &=i\hat{P}_\mu
\end{aligned}
\end{equation*}

If one computes all possible combinations of commutators, one finds the full Conformal Algebra
\begin{equation}
\begin{aligned}             
    [\hat{P}_\mu,\hat{M}_{\nu\rho}]&=i(\eta_{\mu\nu}\hat P_\rho-\eta_{\mu\rho}\hat P_\nu) \\
    [\hat{K}_\rho,\hat{M}_{\mu\nu}]&=i(\eta_{\mu\rho}\hat{K}_\nu-\eta_{\nu\rho}\hat{K}_\mu) \\    [\hat{M}_{\mu\nu},\hat{M}_{\rho\sigma}]&=i(\eta_{\nu\rho}\hat M_{\mu\sigma}-\eta_{\nu\sigma}\hat M_{\mu\rho}-\eta_{\mu\rho}M_{\nu\sigma}+\eta_{\mu\sigma}M_{\nu\rho})  \\
    [\hat{D},\hat{P}_\mu]&=-i\hat{P}_\mu \\
    [\hat{D},\hat{K}_\mu]&=i\hat{K}_\mu \\
    [\hat{P}_\mu,\hat{K}_\nu]&=-2i(\eta_{\mu\nu}\hat{D}+\hat{M}_{\mu\nu})
\end{aligned}
\end{equation}
where all unlisted commutators are zero.

\section{Classifications of Operators in CFT} \label{classification}

As is familiar from Quantum Mechanics, when we quantize our Conformal Field Theory, many of our variables are promoted to operators, which will act on our Hilbert Space. For instance, all of the generators introduced in the previous section will be operators. In fact, the conventions presented in the previous section are useful in part because they guarantee observables, like Momentum, are Hermitian.

There are many ways operators in our CFT will be classified. For instance, they can be classified by their Lorentz Representation (e.g. Momentum $\hat{P}_\mu$ is a Lorentz vector, while Dilitation $\hat{D}$ is a Lorentz scalar). We can also classify operators by their Dimension, which was discussed in section \ref{dimension}. From this, we can consider whether the operator is relevant, irrelevant, or marginal, as discussed in section \ref{rgflow}. This will let us know if perturbing operators will cause the theory to flow away from the conformal point or not. It will also be associated with their eigenvalue under dilatations, as we will discuss below.

An additional classification of operators that is special to CFTs, and will be incredibly important, is as Primary or Decendant Operators. All operators in our CFT will be able to be classified as one or the other.

\subsection{Primaries and Descendants} \label{Primaries}

In order to motivate this new classification, it is helpful to consider an alternative convention for the representation of the generators of conformal transformations. This alternative convention is also especially useful when working in Euclidean signature, as is often done in CFTs. In this convention, the represetations of the transformations are
\begin{equation*}
\begin{aligned}
    \hat{P}_\mu&=\partial_\mu \\
    \hat{M}_{\mu\nu}&=x_\nu\partial_\mu-x_\mu\partial_\nu \\
    \hat{D}&=x^\mu\partial_\mu \\
    \hat{K}_\mu&=2x_\mu x^\rho\partial_\rho-x^2\partial_\mu 
\end{aligned}
\end{equation*}
which lead to the Conformal Algebra
\begin{equation*}
\begin{aligned}
    [\hat{M}_{\mu\nu},\hat{P}_\rho]&=\delta_{\nu\rho}\hat{P}_\mu-\delta_{\mu\rho}\hat{P}_\nu \\
    [\hat{M}_{\mu\nu},\hat{K}_\rho]&=\delta_{\nu\rho}\hat{K}_\mu-\delta_{\mu\rho}\hat{K}_\nu \\
    [\hat{M}_{\mu\nu},\hat{M}_{\rho\sigma}]&=\delta_{\nu\rho}\hat{M}_{\mu\sigma}-\delta_{\mu\rho}\hat{M}_{\nu\sigma}+\delta_{\nu\sigma}\hat{M}_{\rho\mu}-\delta_{\mu\rho}\hat{M}_{\rho\nu} \\
    [\hat{D},\hat{P}_\mu]&=\hat{P}_\mu \\
    [\hat{D},\hat{K}_\mu]&=-\hat{K}_\mu \\
    [\hat{K}_\mu,\hat{P}_\nu]&=2\delta_{\mu\nu}\hat{D}-2\hat{M}_{\mu\nu}
\end{aligned}
\end{equation*}
Again, all other commutators that are not listed vanish. Note that in Euclidean signature $\eta_{\mu\nu}=\delta_{\mu\nu}$. With this convention, we can explicitly see that there is an analogy here with the Quantum Harmonic Oscillator that we know and love from our QM course. In particular, we see that the Momentum operator is analogous to the raising operator of the Harmonic Oscillator, while the Special Conformal operator is analogous to the lowering operator.

\begin{equation*}
\begin{aligned}
    [\hat{D},\hat{P}_\mu]&=\hat{P}_\mu  &  &\longleftrightarrow  & 
 [\hat{N},\hat{a}^\dagger]&=\hat{a}^\dagger\\
    [\hat{D},\hat{K}_\mu]&=-\hat{K}_\mu  &  &\longleftrightarrow  & 
 [\hat{N},\hat{a}]&=-\hat{a}
\end{aligned}
\end{equation*}
where $\hat{a}^\dagger$ and $\hat{a}$ are the raising and lowering operators for the harmonic oscillator, respectively, and $\hat{N}$ is the number operator. Note that the Harmonic Oscillator is reviewed in section \ref{HarmOsc} for your convenience. This striking similarity suggests that $\hat{P}_\mu$ and $\hat{K}_\mu$ can be thought of as the raising and lowering operators of $\hat{D}$, which is indeed the case. Let's see how this works!

We will choose to work in a basis where our states have a definite eigenvalue, $\Delta$, under dilitations. $$|\Delta\rangle= \text{ state with definite dimension } \Delta$$ These states will be created by acting with an operator $\hat{O}_\Delta(0)$ at the origin, on the vacuum 
\begin{equation}
    \hat{O}_\Delta|0\rangle=|\Delta\rangle
\end{equation}
Note that these operators might have a nontrivial Lorentz transformation, but we are suppressing Lorentz indices for notational convenience. You might wonder if it is possible to generate all states in our Hilbert space by acting with a local operator at the origin. Happily, this is possible by the State-Operator Correspondence which is described later, in section \ref{stateoperator}. These operators, $\hat{O}_\Delta$, will satisfy the following commutation relation
\begin{equation} \label{dilcom}
    [\hat{D},\hat{O}_\Delta(0)]=\Delta\hat{O}_\Delta(0)
\end{equation}

We can see that this is all consistent by acting on our state with $\hat D$ to confirm it is an eigenstate

\begin{equation*}
\begin{aligned}
    \hat{D}|\Delta\rangle&=\hat{D}\hat{\mathcal{O}}_\Delta(0)|0\rangle \\
    &=\big ( [\hat{D},\hat{\mathcal{O}}_\Delta(0)]+\hat{\mathcal{O}}_\Delta(0)\hat{D} \big )|0\rangle \\
    &=\Delta\hat{\mathcal{O}}(0)|0\rangle \\
    &=\Delta|\Delta\rangle
\end{aligned}
\end{equation*}
where we used the fact that $\hat D$ is associated with a symmetry of our system and therefore annihilates the vacuum.

We have shown that $|\Delta\rangle$ is in fact an eigenstate of $\hat{D}$ with eigenvalue $\Delta$. Note that $\hat{D}$ commutes with $M_{\mu\nu}$, so these states will also be eigenstates of $\hat{M}_{\mu\nu}$ and we will call their eigenvalues $S_{\mu\nu}$.

In the same way as how the lowering operator lowers the eigenvalue of the number operator for states in the Harmonic Oscillator, the Special Conformal Operator will act like a lowering operator of the Dilitation Operator. It is therefore reasonable to think that, like in the Harmonic Oscillator, we will have a lowest weight state for our states of definite dimension. This is indeed the case and we will call these states Primary States. The operators that act on the vacuum to create these states will be called \textit{Primary Operators}. We will use the notation $\hat{\mathcal{O}}_\Delta$ to denote these, so as to differentiate them from general operators $\hat{O}_\Delta$ shown above. Unlike for the standard Harmonic Oscillator from undergraduate QM, where you only have one lowering operator, in CFTs, we will generally have multiple (some times infinite) Primary (lowest weight) states.

We want these states to be annihilated by $\hat{K}_\mu$. In order for that to happen, the Primary Operators must commute with $\hat K_\mu$. In fact, this is often the definition given of a Primary Operator.
$$[\hat{K}_\mu,\hat{\mathcal{O}}_\Delta(0)]=0$$

We can see that this indeed will mean that the Primary States are annihilated by $\hat K_\mu$
\begin{equation*}
\begin{aligned}
    \hat{K}_\mu|\Delta\rangle&=\hat{K}_\mu\hat{\mathcal{O}}_\Delta(0)|0\rangle \\
    &=\big ( [\hat{K}_\mu,\hat{\mathcal{O}}_\Delta(0)]+\hat{\mathcal{O}}_\Delta(0)\hat{K}_\mu \big ) |0\rangle=0
\end{aligned}
\end{equation*}{}
where we again used the fact that operators associated with symmetries of your theory (in this case, $\hat K$) will annihilate the vacuum. 

If these are our lowest weight states, then what happens when we act on them with momentum, which acts like a raising operator? Let's consider a new state $|\psi\rangle$ that is created by acting on an eigenstate of the dilitation operator.
$$\hat{P}_\mu|\Delta\rangle=|\psi\rangle$$
Is this still an eigenstate of $\hat D$? Let's see...
\begin{equation*}
\begin{aligned}
    \hat{D}|\psi\rangle&=\hat{D}\hat{P}_\mu|\Delta\rangle \\
    &=\big ( [\hat{D},\hat{P}_\mu]+\hat{P}_mu\hat{D} \big )|\Delta\rangle \\
    &=(\hat{P}_\mu+\hat{P}_\mu\hat{D}|\Delta\rangle \\
    &=(\hat{P}_\mu+\hat{P}\Delta)|\Delta\rangle \\
    &=(\Delta+1)\hat{P}_\mu|\Delta\rangle=(\Delta+1)|\psi\rangle
\end{aligned}
\end{equation*}
We see that the state $|\psi\rangle$ is also an eigenstate of $\hat D$, but with dimension $\Delta+1$. As promised, momentum is acting like a raising operator, i.e. $$\hat{P}_\mu|\Delta\rangle=|\Delta+1\rangle$$

If we start with a state created by acting with a Primary Operator $|\Delta\rangle$, and then act with Momentum, we will call this new state a Descendant State. Furthermore, we can consider the operator created when you combine a Primary Operator with Momentum. We will call this a \textit{Descendant Operator}. In fact, we can take this further and consider acting with multiple Momentum Operators to generate more and more descendants. Each time we act with $\hat P_\mu$, the dimension increases by one. If you act with $n$ Momenta, then your operator will have dimension $\Delta+n$. In this way, we can generate an infinite family of operators, called a \textit{Conformal Family}.

To summarize, our \textit{Conformal Family} is the set of operators shown below

\begin{equation*}
\begin{aligned}
    \hat{\mathcal{O}}_\Delta&=\text{Primary operator with dimension }\Delta \\
    \hat{P}_\mu\hat{\mathcal{O}}_\Delta&=\text{Descendant (level 1) with dimension }\Delta+1 \\
    \hat{P}_{\mu_1}...\hat{P}_{\mu_n}\hat{\mathcal{O}}_\Delta&=\text{Descendent (level $n$) with dimension }\Delta+n \\[-1ex]
    &\mathrel{
    \settowidth{\dimen0}{$=$}
    \hbox to \dimen0{\hss$\vdots$\hss}
    }
\end{aligned}
\end{equation*}

 As mentioned previously, you can have multiple Primary Operators in a given CFT. Therefore, in general, you have multiple Conformal Families (oftentimes infinite).

This structure of organizing your operators by Primaries and Descendants is incredibly useful in CFTs. It turns out that the transformation properties of Primary Operators are very nice. Indeed, they transform like tensor densities, which mean they don't transform quite as nicely as a true tensor, but they're the next best thing. An example of a tensor density that you might have seen before is the Levi-Civita symbol $\epsilon_{\mu_1,\mu_2\dots\mu_n}$. Tensor densities transform as follows
\begin{equation} \label{densitytransf}
    \hat{\tilde{\mathcal{O}}}^A_\Delta(\tilde{x}^\mu)=\bigg |\frac{\partial x^\mu}{\partial \tilde{x}^\nu} \bigg |^{\Delta/D}L^A_B\hat{\mathcal{O}}^B_\Delta(x^\mu)
\end{equation}
where $|\frac{\partial x^\mu}{\partial \tilde{x}^\nu}|$ is the Jacobian associated with your transformation and $L^A_B$ is a matrix transforming any Lorentz indices, $A,B,\dots$, associated with the operator.

This fact will be very helpful in section \ref{correlators}, when we derive the form of correlation functions in CFTs. There, we will focus on correlation functions involving only Primary Operators. This is because Descendants will have much more complicated transformation properties. Furthermore, since descendants are created by acting with momentum, which can be represented by a derivative, we can calculate the correlators involving descendants by simply taking derivatives of the correlators with only primaries.

\subsection{Unitarity bounds on spectrum} \label{unitaritybound}
In the discussion thus far, it seems that we are fairly free to add whatever kind of operator we like to our CFT. However, it turns out that we have additional restrictions on what is allowed if we want our theory to represent a system in the real world. In particular, one of the axioms of Quantum Mechanics is that your system must evolve unitarily. That is, your time evolution operator must be a unitary operator ($\hat U^\dagger \hat U = \hat I$). This guarantees that the norms of states are always positive and therefore a probabilistic interpretation of Quantum Mechanics makes sense. Enforcing that the norms of states be positive in our CFT constrains the dimensions of allowed operators. In particular, all operators for a CFT with spacetime dimension $D$ have dimensions constrained by
\begin{align*}
\Delta&=0  & \text{ONLY for the Identity} \\
\Delta&\geq \frac{D-2}{2}  & \text{for scalars} \\
\Delta&\geq\ell+D-2  & \text{for operators with spin } \ell
\end{align*}
Note that in Euclidean theories, enforcing Unitarity is equivalent to enforcing Reflection Positivity.

\section{Correlation functions in CFT} \label{correlators}
In this section, we will derive the form of one-, two-, and three-point functions for ALL Conformal Field Theories. Here is where we will see the real power of symmetry. We will see that the form of these correlation functions, up to a set of constants, is completely determined by symmetry.

Before completing these derivations, let's talk a little about correlation functions, which may be unfamiliar to some readers. A correlation function is simply any object of the form
$$\langle \hat{\mathcal{O}}_1(x_1^\mu)\hat{\mathcal{O}}_2(x_2^\mu)...\hat{\mathcal{O}}_n(x_n^\mu)\rangle=\langle0| \hat{\mathcal{O}}_1(x_1^\mu)\hat{\mathcal{O}}_2(x_2^\mu)...\hat{\mathcal{O}}_n(x_n^\mu)|0\rangle$$
where the $\hat{\mathcal{O}_i}$ are local operators and $|0\rangle$ is the vacuum state.  A correlation function with only one operator is referred to as a one-point function, a correlation function with two operators is a two-point function, and so on.

Correlation functions can have many different physical interpretations depending on what system you are dealing with. If you are studying the Ising model, which describes a system made up of spins on a lattice, you might consider a correlation function with insertions of the spin operator. This then might tell you how correlated the spins are at different places on your lattice. On the other hand, if you are doing particle physics, they might be associated with a scattering amplitude or decay rate. These objects are incredibly important because they are directly related to the things you can measure in a lab. 

Critical to this discussion is the fact that if your system has a particular symmetry, that means your correlation functions should be invariant under the symmetry transformations. Therefore, all correlation functions in a CFT should be invariant under the full set of conformal transformations. Enforcing this is what will allow us to derive the form of the one-, two-, and three-point functions. Explicitly, what this means is that if we consider a transformation $x^\mu \rightarrow \tilde{x}^\mu(x)$, our correlation functions should not change. On the other hand, operators are not conformally invariant (in general, $\hat{\tilde{\mathcal{O}}} \neq \hat{\mathcal{O}}$). As we saw in the previous section, Primary Operators transform as tensor densities (eqn \ref{densitytransf}). Therefore, we want to enforce the following
\begin{equation}
\langle \hat{\mathcal{O}}_1(x_1^\mu)\hat{\mathcal{O}}_2(x_2^\mu)...\hat{\mathcal{O}}_n(x_n^\mu)\rangle=\langle \hat{\tilde{\mathcal{O}}}_1(x_1^\mu)\hat{\tilde{\mathcal{O}}}_2(x_2^\mu)...\hat{\tilde{\mathcal{O}}}_n(x_n^\mu)\rangle
\end{equation}
notice that we have transformed the operators, but we then compare the RHS to the LHS evaluated at the same points. It doesn't make sense to compare the correlation functions when evaluated at different points. This equality must hold regardless of what points you consider. For convenience, we will instead enforce it at the transformed points $\tilde x^\mu$
\begin{equation} \label{corrtransf}
\langle \hat{\mathcal{O}}_1(\tilde x_1^\mu)\hat{\mathcal{O}}_2(\tilde x_2^\mu)...\hat{\mathcal{O}}_n(\tilde x_n^\mu)\rangle=\langle \hat{\tilde{\mathcal{O}}}_1(\tilde x_1^\mu)\hat{\tilde{\mathcal{O}}}_2(\tilde x_2^\mu)...\hat{\tilde{\mathcal{O}}}_n(\tilde x_n^\mu)\rangle
\end{equation}

As discussed in the previous section, all operators in CFT fall into one of two categories, Primaries and Descendants. A Conformal Family consists of one primary operator along with all of its descendants, where descendants are found by applying the momentum operator (i.e. taking the derivative) to the primary. Additionally, Primary Operators have nice transformation properties. With this in mind we will only concern ourselves with Primary Operators, as once we find these it is just a matter of taking derivatives to build the entire family. We will start by looking at just the scalar primaries, as they are the simplest case.

Recall, Primary Operators transform as densities
\begin{equation} \tag{\ref{densitytransf}}
    \hat{\tilde{\mathcal{O}}}^A_\Delta(\tilde{x}^\mu)=\bigg |\frac{\partial x^\mu}{\partial \tilde{x}^\nu} \bigg |^{\Delta/D}L^A_B\hat{\mathcal{O}}^B_\Delta(x^\mu)
\end{equation}
here $|\frac{\partial x^\mu}{\partial \tilde{x}^\nu}|$ is the Jacobian and $L^A_B$ is a matrix transforming any Lorentz indices. Since we will start by considering only scalars, we will not need the $L^A_B$.

Since we will need them, the Jacobians for the conformal transformations are given below. 

\begin{equation*}
\begin{aligned}
    \text{Translation: } \bigg | \frac{\partial x^\mu}{\partial \tilde{x}^\nu} \bigg | &=1 \\
    \text{Rotation: } \bigg | \frac{\partial x^\mu}{\partial \tilde{x}^\nu} \bigg | &=1 \\
    \text{Dilatation: } \bigg | \frac{\partial x^\mu}{\partial \tilde{x}^\nu} \bigg | &=\lambda^{-D} \\
    \text{Inversion: } \bigg | \frac{\partial x^\mu}{\partial \tilde{x}^\nu} \bigg | &=\bigg ( \frac{1}{\tilde{x}^2} \bigg )^D 
\end{aligned}
\end{equation*}
We will not derive all of these. We will show the derivation for the dilitation operator, but the others are left as an exercise for the reader.

First, let's rewrite the dilatation transformation as $$x^\mu=\frac{\tilde{x}^\mu}{\lambda}$$
Now, to get the Jacobian we must take the partial derivative of this with respect to $\tilde{x}^\nu$
$$\frac{\partial x^\mu}{\partial \tilde{x}^\nu}=\frac{\delta^\mu_\nu}{\lambda}$$ 
We then take the determinant of this, in which case we will get a factor of $\frac{1}{\lambda}$ for each of our spacetime dimensions. So our Jacobian for dilatation is $$\bigg | \frac{\partial x^\mu}{\partial \tilde{x}^\nu}\bigg |=\bigg |\frac{\delta^\mu_\nu}{\lambda}\bigg |=\lambda^{-D}$$
as shown above.

\subsection{One-point Function of a Scalar Primary}

A scalar primary is denoted by $\hat{\mathcal{O}}_\Delta$ (notice the absence of the Lorentz index, which indicates that it is a scalar operator). We want to enforce eqn \ref{corrtransf}. For a one-point function, this reduces to 
\begin{equation} \label{onepointcond}
    \langle \hat{\mathcal{O}}_\Delta(\tilde{x}^\mu) \rangle = \langle \hat{\tilde{\mathcal{O}}}_\Delta(\tilde{x}^\mu) \rangle
\end{equation}
This condition must be enforced for all four conformal transformations. We will begin by enforcing translation.
\vspace{5mm}

\noindent \underline{Translation}: $\tilde{x}^\mu=x^\mu+a^\mu$

It was given previously that the Jacobian for a translation is one. Therefore, our operator simply does not change under this translation $$\hat{\tilde{\mathcal{O}}}_\Delta(\tilde{x}^\mu)=\hat{\mathcal{O}}_\Delta(x^\mu)$$
enforcing this in (\ref{onepointcond}), we are left with $$\langle \hat{\mathcal{O}}_\Delta(\tilde{x}^\mu) \rangle=\langle \hat{\mathcal{O}}_\Delta(x^\mu) \rangle$$
where the operators are now the same on both sides of the equation. Notice that a correlation function is just a function. Therefore, this is equivalent to saying $$f(\tilde{x}^\mu)=f(x^\mu)$$ Since this must be true for every possible translation, this tells us that the function has the same output regardless of what the input is, which means the function must just be some constant. Therefore, by enforcing translation we can conclude that $$\langle \hat{\mathcal{O}}_\Delta(\tilde{x}^\mu) \rangle=\langle \hat{\mathcal{O}}_\Delta(x^\mu) \rangle=\text{constant}=C$$

We are not yet done. We need to make sure that all four transformations leave the one-point function invariant. Let's see what we can learn when we enforce dilatation.

\hspace{1cm}

\noindent \underline{Dilatation}: $\tilde{x}^\mu=\lambda x^\mu$ 

Applying the Jacobian for dilitation, we see our Primary Scalar Operator transforms as $$\hat{\tilde{\mathcal{O}}}_\Delta(\tilde{x}^\mu)=\bigg |\frac{\partial x^\mu}{\partial \tilde{x}^\nu}\bigg|^{\Delta / D}\hat{\mathcal{O}}_\Delta(x^\mu)=\lambda^{-\Delta}\hat{\mathcal{O}}_\Delta(x^\mu)$$

We want to enforce this in (\ref{onepointcond}) and use our results from enforcing translation invariance. This gives us
\begin{equation*}
\begin{aligned}
    \langle \hat{\mathcal{O}}_\Delta(\tilde{x}^\mu) \rangle&=\langle \hat{\tilde{\mathcal{O}}}_\Delta(\tilde{x}^\mu) \rangle \\
    &=\langle \lambda^{-\Delta} \hat{\mathcal{O}}_\Delta (x^\mu) \rangle \\
    &=\lambda^{-\Delta}\langle \hat{\mathcal{O}}_\Delta (x^\mu) \rangle 
\end{aligned}
\end{equation*}

We found previously, by enforcing translation, that $$\langle \hat{\mathcal{O}}_\Delta(\tilde{x}^\mu) \rangle=C$$
which means $$C=\lambda^{-\Delta}C$$
This equation must be true for arbitrary scale factor $\lambda$. Therefore, unless $\Delta=0$, we can conclude that $C=0$.

For unitary CFTs, the only $\Delta=0$ operator is the identity operator. So, with the exception of the identity, all one-point functions must vanish! 
\begin{equation}
    \boxed{\langle \hat{\mathcal{O}}_\Delta(x^\mu) \rangle=0 \text{ for } \Delta \neq 0}
\end{equation}

We said that we must impose all four conformal transformations, but the others are trivially satisfied at this point. So we are done with one-point correlators! Again, we'd like to highlight the fact that this is the result for ALL CFTs. You don't need to know anything else about the system, only that it has conformal symmetry.

\subsection{Two-point Scalar Primary}

For two-point functions, we need to enforce

\begin{equation} \label{twopointtrans}
    \langle \hat{\mathcal{O}}_{\Delta_1}(\tilde{x}_1^\mu)\hat{\mathcal{O}}_{\Delta_2}(\tilde{x}_2^\mu)\rangle=\langle \hat{\tilde{\mathcal{O}}}_{\Delta_1}(\tilde{x}_1^\mu)\hat{\tilde{\mathcal{O}}}_{\Delta_2}(\tilde{x}_2^\mu)\rangle
\end{equation}
Again, this must be done for all four conformal transformations. As with the one-point function, we will begin by enforcing translation.

\hspace{1cm}

\noindent \underline{Translation}:

First, notice that $\langle \hat{\mathcal{O}}_{\Delta_1}(\tilde{x}_1^\mu)\hat{\mathcal{O}}_{\Delta_2}(\tilde{x}_2^\mu)\rangle$ is an object that takes two positions as inputs and gives back a number, so we can just write this as a function of $\tilde{x}_1^\mu$ and $\tilde{x}_2^\mu$ $$\langle \hat{\mathcal{O}}_{\Delta_1}(\tilde{x}_1^\mu)\hat{\mathcal{O}}_{\Delta_2}(\tilde{x}_2^\mu)\rangle=f(\tilde{x}_1^\mu,\tilde{x}_2^\mu)$$
We found previously that under translations, scalar primary operators transform as $$\hat{\tilde{\mathcal{O}}}(\tilde{x}^\mu)=\hat{\mathcal{O}}(x^\mu)$$
Putting this result in the right side of equation (\ref{twopointtrans}), we find that $$\langle \hat{\mathcal{O}}_{\Delta_1}(\tilde{x}_1^\mu)\hat{\mathcal{O}}_{\Delta_2}(\tilde{x}_2^\mu)\rangle=\langle \hat{\mathcal{O}}_{\Delta_1}(x_1^\mu)\hat{\mathcal{O}}_{\Delta_2}(x_2^\mu)\rangle$$
Notice that the only differences between the left and right side of this equation are the inputs. The function on each side is the same $$f(\tilde{x}_1^\mu,\tilde{x}_2^\mu)=f(x_1^\mu,x_2^\mu)$$ 
Under translation, we have $\tilde{x}^\mu=x^\mu+a^\mu \rightarrow x^\mu=\tilde{x}^\mu-a^\mu$. If we put this into the previous equation it becomes $$f(x^\mu_1-a^\mu,x^\mu_2-a^\mu)=f(x^\mu_1,x^\mu_2) \text{ , }\forall a^\mu$$
This must be true regardless of the value of $a^\mu$ which means that $a^\mu$ must somehow cancel out. This is only satisfied if it is a function of $x_1^\mu-x_2^\mu$ so we have$$f(x^\mu_1,x^\mu_2)=f(x^\mu_1-x^\mu_2)$$
That is, our function cannot be any function of the two positions. Rather, it can only depend on the displacement between the two positions. Therefore,
$$\langle \hat{\mathcal{O}}_{\Delta_1}(x_1^\mu)\hat{\mathcal{O}}_{\Delta_2}(x_2^\mu) \rangle=f(x^\mu_1-x^\mu_2)$$
Let's now see what we can learn by enforcing rotation.

\hspace{1cm}

\noindent \underline{Rotation} $\tilde{x}^\mu=\Lambda^\mu_\nu x^\mu$

The Jacobian for rotation is the same as for translation, 1. Therefore, scalar primary operators transform the same under rotation as they do under translation $$\langle \hat{\mathcal{O}}_{\Delta_1}(\tilde{x}_1^\mu)\hat{\mathcal{O}}_{\Delta_2}(\tilde{x}_2^\mu)\rangle=\langle \hat{\mathcal{O}}_{\Delta_1}(x_1^\mu)\hat{\mathcal{O}}_{\Delta_2}(x_2^\mu)\rangle$$
Re-expressing this in a more familiar form, as functions, we have $$f(\tilde{x}_1^\mu,\tilde{x}_2^\mu)=f(x_1^\mu,x_2^\mu)$$
Next, we impose what we found by imposing translational invariance $$f(\tilde{x}_1^\mu-\tilde{x}_2^\mu)=f(x_1^\mu-x_2^\mu)$$
Expressing the transformed coordinates in terms of our original coordinate system, we find $$f(\Lambda^\mu_\nu(x^\nu_1-x^\nu_2))=f(x^\mu_1-x^\mu_2)$$
This tells us that applying a rotation has no effect on the output. This means that the function must depend only on the magnitude of the separation $|x_1^\mu-x_2^\mu|$ (Recall, rotating a vector changes it, but rotating a scalar does nothing). So, from applying translational and rotational invariance, we can conclude that

$$\langle \hat{\mathcal{O}}_{\Delta_1}(x_1^\mu)\hat{\mathcal{O}}_{\Delta_2}(x_2^\mu)\rangle=f(|x^\mu_1-x^\mu_2|)$$
We will now continue by enforcing invariance under dilatation.

\hspace{1cm}

\noindent \underline{Dilatation}

Recall, under dilatation, scalar primary operators transform as $$\hat{\tilde{\mathcal{O}}}_\Delta(\tilde{x}^\mu)=\lambda^{-\Delta}\hat{\mathcal{O}}_\Delta(x^\mu)$$ 
Substituting this into our two-point function condition, eqn. (\ref{twopointtrans}), we have
\begin{equation*}
\begin{aligned}
    \langle \hat{\mathcal{O}}_{\Delta_1}(\tilde{x}_1^\mu)\hat{\mathcal{O}}_{\Delta_2}(\tilde{x}_2^\mu)\rangle&=\langle \hat{\tilde{\mathcal{O}}}_{\Delta_1}(\tilde{x}_1^\mu)\hat{\tilde{\mathcal{O}}}_{\Delta_2}(\tilde{x}_2^\mu)\rangle \\
    &=\langle \lambda^{-\Delta_1}\hat{\mathcal{O}}_{\Delta_1}(x_1^\mu)\lambda^{-\Delta_2}\hat{\mathcal{O}}_{\Delta_2}(x_2^\mu)\rangle \\
    &=\lambda^{-\Delta_1}\lambda^{-\Delta_2}\langle \hat{\mathcal{O}}_{\Delta_1}(x_1^\mu)\hat{\mathcal{O}}_{\Delta_2}(x_2^\mu)\rangle \\
    &=\lambda^{-(\Delta_1+\Delta_2)}f(|x_1^\mu-x_2^\mu|)
\end{aligned}
\end{equation*}
where we are able to pull the $\lambda$'s out of the correlator because they are just scalars. Notice also that we used what we already learned from translational and rotational invariance. This tells us that $$f(|\tilde{x}_1^\mu-\tilde{x}_2^\mu|)=\lambda^{-(\Delta_1+\Delta_2)}f(|x_1^\mu-x_2^\mu|)$$
We can apply the transformation to the coordinates on the left-hand-side, which gives
\begin{equation*}
    f(\lambda|x_1^\mu-x_2^\mu|)=\lambda^{-(\Delta_1+\Delta_2)}f(|x_1^\mu-x_2^\mu|)
\end{equation*}
What does this mean? We can consider expanding our function in a power series
$$f(|x_1^\mu-x_2^\mu|)=\sum_n c_n|x_1^\mu-x_2^\mu|^n$$
Substituting this in above gives
\begin{equation*}
    \sum_n c_n\lambda^n|x_1^\mu-x_2^\mu|^n=\lambda^{-(\Delta_1+\Delta_2)}\sum_n c_n|x_1^\mu-x_2^\mu|^n
\end{equation*}
This is only satisfied for all $\lambda$, if all $n=0$ except $n=-(\Delta_1+\Delta_2)$.
Therefore, after enforcing translation, rotation, and dilatation symmetry we have
\begin{equation}
\langle \hat{\mathcal{O}}_{\Delta_1}(x_1^\mu)\hat{\mathcal{O}}_{\Delta_2}(x_2^\mu)\rangle=C|x_1^\mu-x_2^\mu|^{-(\Delta_1+\Delta_2)}
\end{equation}
where $C$ is some undetermined constant.

\hspace{1cm}

\noindent \underline{Special Conformal Transformation}

Enforcing special conformal symmetry directly is a very messy business. Luckily for us, as discussed previously, a special conformal transformation is equivalent to performing an inversion, followed by a translation, followed by another inversion. Since we have already enforced translational invariance, this means it is sufficient to enforce inversion invariance, which is much easier. Recall, an inversion is given by $$x^\mu=\frac{\tilde{x}^\mu}{\tilde{x}^2}$$

As with the other transformations, we need the Jacobian for inversion in order to see how the operators will transform. This is given by

\begin{equation*}
    \bigg | \frac{\partial x^\mu}{\partial \tilde{x}^\nu} \bigg |= \frac{1}{\tilde{x}^{2D}}
\end{equation*}

Therefore, under inversion, scalar primary operators transform as $$\hat{\tilde{\mathcal{O}}}_\Delta(\tilde{x}^\mu)=\bigg ( \frac{1}{\tilde{x}^{2D}} \bigg )^{\Delta/D}\hat{\mathcal{O}}_\Delta(x^\mu)=\frac{1}{(\tilde{x}^2)^\Delta}\hat{\mathcal{O}}_\Delta(x^\mu)$$
As usual, we will now go put this into equation (\ref{twopointtrans}) to enforce the symmetry
\begin{equation*}
\begin{aligned}
    \langle \hat{\mathcal{O}}_{\Delta_1}(\tilde{x}_1^\mu)\hat{\mathcal{O}}_{\Delta_2}(\tilde{x}_2^\mu)\rangle &=\langle \hat{\tilde{\mathcal{O}}}_{\Delta_1}(\tilde{x}_1^\mu)\hat{\tilde{\mathcal{O}}}_{\Delta_2}(\tilde{x}_2^\mu)\rangle \\
    &=\bigg \langle \frac{1}{(\tilde{x}_1^2)^{\Delta_1}}\hat{\mathcal{O}}_{\Delta_1}(x_1^\mu)\frac{1}{(\tilde{x}_2^2)^{\Delta_2}}\hat{\mathcal{O}}_{\Delta_2}(x_2^\mu)\bigg \rangle \\
    &=\frac{1}{(\tilde{x}_1^2)^{\Delta_1}}\frac{1}{(\tilde{x}_2^2)^{\Delta_2}}\langle\hat{\mathcal{O}}_{\Delta_1}(x_1^\mu)\hat{\mathcal{O}}_{\Delta_2}(x_2^\mu)\rangle
\end{aligned}
\end{equation*}
Now, we can use our result from enforcing dilitation to replace $\langle \hat{\mathcal{O}}_{\Delta_1}(\tilde{x}_1^\mu)\hat{\mathcal{O}}_{\Delta_2}(\tilde{x}_2^\mu)\rangle$ on the left and $\langle\hat{\mathcal{O}}_{\Delta_1}(x_1^\mu)\hat{\mathcal{O}}_{\Delta_2}(x_2^\mu)\rangle$ on the right of this equation to get 
\begin{equation} 
    \frac{C}{|\tilde{x}_1^\mu-\tilde{x}_2^\mu|^{\Delta_1+\Delta_2}}=\frac{1}{(\tilde{x}_1^2)^{\Delta_1}}\frac{1}{(\tilde{x}_2^2)^{\Delta_2}}\frac{C}{|x_1^\mu-x_2^\mu|^{\Delta_1+\Delta_2}}
\end{equation}
With a bit of algebra, this is equivalent to 
\begin{equation} \label{twopointinvtrans}
    \frac{(\tilde{x}_1^2)^{\Delta_1}(\tilde{x}_2^2)^{\Delta_2}}{|\tilde{x}_1^\mu-\tilde{x}_2^\mu|^{\Delta_1+\Delta_2}}=\frac{1}{|x_1^\mu-x_2^\mu|^{\Delta_1+\Delta_2}}
\end{equation}
In order to put this in a more friendly form, we will use the following identity for inversions. Note: verifying this relationship requires substituting in the inversion transformation and some algebra. The reader is highly encouraged to check it. 
\begin{equation}
    \frac{\tilde{x}_1^2\tilde{x}_2^2}{(\tilde{x}_1^\mu-\tilde{x}_2^\mu)^2}=\frac{1}{(x_1^\mu-x_2^\mu)^2}
\end{equation}
Using this identity in eqn. (\ref{twopointinvtrans}), we find 
$$\frac{(\tilde{x}_1^2)^{\Delta_1}(\tilde{x}_2^2)^{\Delta_2}}{|\tilde{x}_1^\mu-\tilde{x}_2^\mu|^{\Delta_1+\Delta_2}}= \bigg [ \frac{\tilde{x}_1^2\tilde{x}_2^2}{|\tilde{x}_1^\mu-\tilde{x}_2^\mu|^2} \bigg ]^{\frac{\Delta_1+\Delta_2}{2}}$$
This is only satisfied if $$\Delta_1=\Delta_2$$ Therefore, we find that the two-point function vanishes, unless the dimensions of the two operators are the same. In summary, the two-point function for scalar primaries in ANY CFT is given by
\begin{equation} \label{twopoint}
    \boxed{\langle \hat{\mathcal{O}}_{\Delta_1}(x_1^\mu)\hat{\mathcal{O}}_{\Delta_2}(x_2^\mu)\rangle=\frac{C\delta_{\Delta_1\Delta_2}}{|x_1^\mu-x_2^\mu|^{\Delta_1+\Delta_2}}}
\end{equation}
Note that it is standard convention to choose to normalize your operators so that $C=1$, so you will often see this without the $C$ constant included. We leave it here for complete generality.

\subsection{Three-point Scalar Primary} \label{threepoint}

For the three-point function, we need to enforce
\begin{equation} \label{threepointtrans}
    \langle \hat{\mathcal{O}}_1(x_1^\mu)\hat{\mathcal{O}}_2(x_2^\mu)\hat{\mathcal{O}}_3(x_3^\mu)\rangle=\langle \hat{\tilde{\mathcal{O}}}_1(x_1^\mu)\hat{\tilde{\mathcal{O}}}_2(x_2^\mu)\hat{\tilde{\mathcal{O}}}_3(x_3^\mu)\rangle
\end{equation}
Enforcing the symmetries for the three-point function follows in a very similar way to the two-point function, so we will not include as much detail. The reader is encouraged to work through any excluded details on their own.

\hspace{1cm}

\noindent \underline{Poincar\'e}

For translations and rotations, the same line of argumentation that was used for two-point functions can be applied. However, instead of two points at our disposal, we have three. Therefore, our function can be a function of the magnitude of the separations between any parings of three points.  $$\langle \hat{\mathcal{O}}_1(x_1^\mu)\hat{\mathcal{O}}_2(x_2^\mu)\hat{\mathcal{O}}_3(x_3^\mu)\rangle=f(|x_{12}^\mu|,|x_{23}^\mu|,|x_{31}^\mu|)$$
where $|x_{12}^\mu|=|x_1^\mu-x_2^\mu|$, $|x_{23}^\mu|=|x_2^\mu-x_3^\mu|$, and $|x_{31}^\mu|=|x_3^\mu-x_1^\mu|$.

\hspace{1cm}

\noindent \underline{Dilatation}

Enforcing dilitation invariance with our three-point function, we have
\begin{equation*}
\begin{aligned}
    \langle \hat{\mathcal{O}}_{\Delta_1}(\tilde{x}_1^\mu)\hat{\mathcal{O}}_{\Delta_2}(\tilde{x}_2^\mu)\hat{\mathcal{O}}_{\Delta_3}(\tilde{x}_3^\mu)\rangle&=\langle \hat{\tilde{\mathcal{O}}}_{\Delta_1}(\tilde{x}_1^\mu)\hat{\tilde{\mathcal{O}}}_{\Delta_2}(\tilde{x}_2^\mu)\hat{\tilde{\mathcal{O}}}_{\Delta_3}(\tilde{x}_3^\mu)\rangle \\
    &=\langle \lambda^{-\Delta_1}\hat{\mathcal{O}}_{\Delta_1}(x_1^\mu)\lambda^{-\Delta_2}\hat{\mathcal{O}}_{\Delta_2}(x_2^\mu)\lambda^{-\Delta_3}\hat{\mathcal{O}}_{\Delta_3}(x_3^\mu)\rangle \\
    &=\lambda^{-\Delta_1}\lambda^{-\Delta_2}\lambda^{-\Delta_3}\langle \hat{\mathcal{O}}_{\Delta_1}(x_1^\mu)\hat{\mathcal{O}}_{\Delta_2}(x_2^\mu)\hat{\mathcal{O}}_{\Delta_3}(x_3^\mu)\rangle 
\end{aligned}
\end{equation*}
Using our results from enforcing Poincar\'e invariance, this becomes
\begin{equation}
    f(|\tilde{x}_{12}^\mu|,|\tilde{x}_{23}^\mu|,|\tilde{x}_{31}^\mu|)=\lambda^{-(\Delta_1+\Delta_2+\Delta_3)}f(|x_{12}^\mu|,|x_{23}^\mu|,|x_{31}^\mu|)
\end{equation}
Subsituting in the dilitation transformation on the LHS, this is
\begin{equation}
    f(\lambda|x_{12}^\mu|,\lambda|x_{23}^\mu|,\lambda|x_{31}^\mu|)=\lambda^{-(\Delta_1+\Delta_2+\Delta_3)}f(|x_{12}^\mu|,|x_{23}^\mu|,|x_{31}^\mu|)
\end{equation}
As with the two-point function, we can expand our function in a power series.
\begin{equation}
    f(|x_{12}^\mu|,|x_{23}^\mu|,|x_{31}^\mu|)=\sum_{nmp}c_{nmp}|x_{12}^\mu|^n|x_{23}^\mu|^m|x_{31}^\mu|^p
\end{equation}
Substituting this in, we find that all terms must vanish, unless $$n+m+p=-(\Delta_1+\Delta_2+\Delta_3)$$ Therefore, dilitation and Poincar\'e invariance tell us
\begin{equation}
    \langle \hat{\mathcal{O}}_1(x_1^\mu)\hat{\mathcal{O}}_2(x_2^\mu)\hat{\mathcal{O}}_3(x_3^\mu)\rangle=\sum_{nmp=-(\Delta_1+\Delta_2+\Delta_3)}c_{nmp}|x_{12}^\mu|_n|x_{23}^\mu|^m|x_{31}^\mu|^p
\end{equation}

\hspace{1cm}

\noindent \underline{Special Conformal Transformation}

Again, to find the effect of imposing special conformal symmetry, we need only to impose inversion symmetry, which is much easier. Although easier, the algebra is still quite nasty and will not be shown here. Ultimately, inversion (therefore special conformal) invariance leads to the additional constraint that  all terms vanish, unless

\begin{equation} \label{nmp}
\begin{aligned}
    n&=\Delta_1+\Delta_2-\Delta_3 \\
    m&=\Delta_1+\Delta_3-\Delta_2 \\
    p&=\Delta_2+\Delta_3-\Delta_1
\end{aligned}
\end{equation}
Therefore, after enforcing all of the conformal symmetries on the 3-point function of scalar primaries, we find
\begin{equation}
    \boxed{\hat{\mathcal{O}}_{\Delta_1}(x_1^\mu)\hat{\mathcal{O}}_{\Delta_2}(x_2^\mu)\hat{\mathcal{O}}_{\Delta_3}(x_3^\mu)=\frac{C_{123}}{|x_{12}^\mu|^n|x_{23}^\mu|^m|x_{31}^\mu|^p}}
\end{equation}
where $n$, $m$, and $p$ are given by (\ref{nmp}). We find that, as was the case with the two-point scalar primaries, the spatial dependence of 3-point scalar primaries are completely determined. We are left only with a set of constants $C_{123}$. It turns out that this set of constants is vitally important to defining any particular conformal field theory and they tell you how much your given operators interact. This set of constants goes by various names including the \textit{3-point coefficients}, the \textit{OPE coefficients}, and the \textit{structure constants}.

\subsection{Four-point Scalar Primary}

Four-point functions are no longer completely determined because once you have at least four points you can create conformally invariant cross-ratios. These are functions of the points that are invariant under all conformal transformations. For four points, we have two independent cross-ratios

\begin{align}
    \eta_1=\frac{|x_{12}^\mu||x_{34}^\mu|}{|x_{13}^\mu||x_{24}^\mu|} && \text{and} &&
    \eta_2=\frac{|x_{12}^\mu||x_{34}^\mu|}{|x_{23}^\mu||x_{14}^\mu|}
\end{align}
As these functions only depend on the magnitude of the separations between points, Poincar\'e invariance is immediately satisfied. The reader is encouraged to verify that these are invariant under dilitations and inversion. Note, that for inversion, the identity we used for the two-point function is helpful.

Since we have these two cross section ratios that are independently conformally invariant, this means we can write any function of these cross sections, $F(\eta_1,\eta_2)$, and this function will also be conformally invariant. To reiterate, this means that 4-point functions \textit{cannot be completely determined} by enforcing the conformal symmetries. If we enforce the conformal symmetries, for a correlation functions of operators all with dimension $\Delta$, we find
\begin{equation}
    \boxed{\langle \hat{\mathcal{O}}_{\Delta}(x_1^\mu)\hat{\mathcal{O}}_{\Delta}(x_2^\mu)\hat{\mathcal{O}}_{\Delta}(x_3^\mu)\hat{\mathcal{O}}_{\Delta}(x_4^\mu)\rangle=F(\eta_1,\eta_2)|x_{12}|^{-2\Delta}|x_{34}|^{-2\Delta}}
\end{equation}
where $F(\eta_1,\eta_2)$ is an undetermined function of the cross-ratios. 

It turns out there is more that can be done. One can add additional constraints on the four-point function by considering, for example, the behavior of the Operator Product Expansion, which is discussed in section \ref{OPE}. This is a big part of what is done in the Conformal Bootstrap Program, which is discussed in section \ref{bootstrap}.

\subsection{Going Beyond Scalars}
The correlation functions we have considered thus far have only involved scalar operators. Of course, we generally will also want to know about correlation functions for operators with non-trivial Lorentz indices. In order to enforce conformal invariance for these objects, one needs to consider the more general Primary Operator transformation
\begin{equation} \tag{\ref{densitytransf}}
    \hat{\tilde{\mathcal{O}}}^A_\Delta(\tilde{x}^\mu)=\bigg |\frac{\partial x^\mu}{\partial \tilde{x}^\nu} \bigg |^{\Delta/D}L^A_B\hat{\mathcal{O}}^B_\Delta(x^\mu)
\end{equation}
where $L^A_B$ is a matrix that accounts for the transformation of the Lorentz indices. As was the case with scalar primaries, one can show that one-, two-, and three-point functions are completely determined up to a set of constants. Again, four-point functions and higher have conformal cross-ratios that allow for a function that is undetermined by conformal invariance alone. 

Let's consider vector operators $\hat{\mathcal{O}}_\Delta^\rho(x^\mu)$. We first need to apply (\ref{densitytransf}) to our primary vector operator.
\begin{equation}     \hat{\tilde{\mathcal{O}}}^\rho_\Delta(\tilde{x}^\mu)=\bigg |\frac{\partial x^\mu}{\partial \tilde{x}^\nu} \bigg |^{\Delta/D} \frac{\partial \tilde{x}^\rho}{\partial x^\sigma} \hat{\mathcal{O}}^\sigma_\Delta(x^\mu)
\end{equation}
As was the case before, one-point functions will vanish.
\begin{equation}
    \boxed{\langle \hat{\mathcal{O}}_\Delta^\rho(x^\mu) \rangle=0 }
\end{equation}
By applying a similar procedure as we did with scalars, but utilizing the transformation for vector densities, you can find the two point function.
\begin{equation}
    \boxed{\langle \hat{\mathcal{O}}_{\Delta_1}^\mu(x) \hat{\mathcal{O}}_{\Delta_2}^\nu(x) \rangle=\frac{C\delta_{\Delta_1\Delta_2}}{|x_1^\mu-x_2^\mu|^{\Delta_1+\Delta_2}}I^{\mu\nu}(x_1-x_2)}
\end{equation}
where you might notice this is exactly the same as the answer for the two-point scalar correlator, except for the $I^{\mu\nu}$. This is necessary for the Lorentz indices to agree and is given by 
\begin{equation} \label{itensor}
    I^{\mu\nu}(x_{12})=\delta^{\mu\nu}-\frac{2x_{12}^\mu x_{12}^\nu}{x_{12}^2}
\end{equation}
where $x_{12}^\mu=x_1^\mu-x_2^\mu$.

One can go further, and compute the three-point function for vector primaries. Just like before, once you have four-points, you can compute invariant cross-ratios, which allow for a function that is undetermined by enforcing symmetry alone.

Similarly, one can compute one-, two-, and three-point functions for higher rank tensors. The computations get increasingly difficult, as for each additional Lorentz index, you need an additional Jacobian matrix in your transformation. However, conceptually, these calculations proceed in the exact same way as they did for the scalar primary correlation fucntions.

\subsection{Going Beyond Primaries}

Once we have the correlation functions for primaries, we only need to take derivatives in order to derive the expressions for decendants. Recall, we can generate a full tower of decendants by acting with the momentum operator, which can be represented as a derivative operator. Let's look explicitly at how this would work for a two-point function. We know that the two-point function for scalar primaries is given by eqn. (\ref{twopoint}), repeated here for convenience.
\begin{equation} \tag{\ref{twopoint}}
    \langle \hat{\mathcal{O}}_{\Delta_1}(x_1^\mu)\hat{\mathcal{O}}_{\Delta_2}(x_2^\mu)\rangle=\frac{C\delta_{\Delta_1\Delta_2}}{|x_1^\mu-x_2^\mu|^{\Delta_1+\Delta_2}}
\end{equation}
Let's say we want to compute the two point function between a primary $\hat{\mathcal{O}}_{\Delta}$ and its first descendant $\hat{P^\nu}\hat{\mathcal{O}}_{\Delta}$. The descendant can be expressed in terms of the derivative operator as $-i\partial^\nu\hat{\mathcal{O}}_{\Delta}$. Therefore, the two-point function is given by
\begin{align} 
    \langle \hat{P}^\nu \hat{\mathcal{O}}_{\Delta}(x_1^\mu)\hat{\mathcal{O}}_{\Delta}(x_2^\mu)\rangle&=-i\frac{\partial}{\partial x_{1 \nu}}\frac{C}{|x_1^\mu-x_2^\mu|^{2\Delta}}\\
    &=\frac{2i\Delta C(x_1^{\nu}-x_2^{\nu})}{|x_1^{\mu}-x_2^\mu|^{2\Delta+2}}
\end{align}
As you can see, finding the correlation functions involving descendants simply requires differentiation.

\section{Further topics in CFT}
This section provides brief discussion on further topics in CFT. The discussion will not go into many technical details as those are beyond the scope of this primer. The goal of this section is to expose the reader to terms and ideas that are used in CFT research for further study.

\subsection{What's so special about 2D?}
In this primer, we have repeatedly mentioned explicitly that we are working in $D\geq 3$. Why have we needed to make this comment? It turns out that 2D CFTs are very special. We have seen that there is a lot of symmetry in CFTs. They are invariant under translations, rotations/boosts, dilatations, and special conformal transformations. We saw that we could associate a generator with each of these ($P_\mu,M_{\mu\nu}, D, K_\mu$). By enforcing these symmetries, we made a lot of progress in solving for universal properties of all CFTs. 

In 2D, something magical happens! When you only have two dimensions, it becomes natural (and very helpful) to work in the complex plain, with $z=x^1+ix^2$ and $\bar z=x^1-ix^2$. If you do this, the Conformal Killing Equation (\ref{confkilling}) in 2D is equivalent to the Cauchy-Riemann equations: $\partial_1\xi_1=\partial_2\xi_2$ and $\partial_1\xi_2=-\partial_2\xi_1$. The solutions to these equations are any analytic coordinate transformations: $z\rightarrow f(z)$ and $\bar z\rightarrow f(\bar z)$. This means that the function defining your transformation for the $z$ coordinate is ANY function of only $z$ and the function defining the transformation for the $\bar z$ coordinate is ANY function of only $\bar z$. If we want to organize this set of transformations into generators, like we did for $D\geq 3$, then instead of only four generators, we have an infinite number. This is because we need to be able to generate ANY analytic function. Therefore, we see the power of symmetry in 2D is greatly enhanced! 

It is important to note that many resources in CFT deal specifically with 2D CFT. There are many excellent reviews of 2D CFT, including \cite{ginsparg1988applied,cardy2008conformal,Yin:2018DH}. One must always be sure to note whether the particular paper/book is dealing with 2D or higher dimensions. Apart from this section, this primer is specifically in $D\geq3$, though the authors are working on a follow up, which will serve as an introduction to 2D CFT for undergraduates.

\subsection{The Stress Energy Tensor}
The Stress Energy Tensor $T^{\mu\nu}$ (also known as the Energy-Momentum Tensor or simply the Stress Tensor) is a very important object in Conformal Field Theory. In every CFT you have, at a minimum, the identity operator and the Stress Tensor.

You may be familiar with stress tensors from other areas of Physics. This is an object that describes the energy-like aspects of a system. Students often first encounter it when learning Special Relativity. There, you learn that the $T^{00}$ component describes the energy density of your system, the $T^{i0}$ and $T^{0i}$ components describe momentum densities and energy fluxes, the $T^{ii}$ components describe pressures, and the off-diagonal $T^{ij}$ components describe shear stresses and momentum fluxes. 

If you go further and study General Relativity, you will learn that this object is the source of the curvature of spacetime. The Einstein equation relates mass and energy to the curvature of spacetime. One side of the equation includes the stress tensor, which acts as a source  for the other side, which describes the geometry of spacetime.

There are two important properties of the stress tensor in Lorentz invariant theories (e.g. Special Relativity and Quantum Field Theory). The first is that it is conserved, i.e. $\partial_\mu T^{\mu\nu}=0$. This tells us that Energy and Momentum are conserved, as you would expect in any Lorentz-invariant theory. The second imporatnt property is that it can be made to be symmetric, i.e. $T^{\mu\nu}=T^{\nu\mu}$. We say ``it can be made to be symmetric," rather than ``it is symmetric" intentionally. That is because there is ambigiuty in the way the stress tensor is defined. This should not be surprising. Recall, for instance, there is always ambiguity in the way you define potential energy. It is only the change in potential energy that is physically measurable and therefore unambiguous. The most accurate statement is therefore that you can always add a correction term to the stress tensor to make it symmetric (this correction term does not change anything physically measurable).

In CFT, there is a third important property: The stress tensor can be made to be traceless ($T^\mu_{\;\mu}=0$). This is true for classical and quantum CFTs. However, as mentioned previously, if you start with a classical CFT and then quantize, sometimes the conformal symmetry will be broken. You can track this by considering the tracelessness property of the stress tensor. If you start with a classical CFT and then quantize, you may find that the stress tensor picks up a non-zero trace. When this happens, it is known as a conformal anomaly (a.k.a. trace anomaly or Weyl anomaly).

In $D\geq 3$, the stress tensor is a primary operator\footnote{Note that in D=2, the stress tensor is a quasi-primary and is actually a descendant of the identity operator.} with dimension $\Delta=D$. If you derive the two-point function following a similar method as was presented in section \ref{correlators}, then you will find
\begin{equation*}
    \langle T_{\mu\nu}(x_1^\delta) T_{\rho\sigma}(x_2^\delta)\rangle = \frac{c}{|x_1^\delta-x_2^\delta|^{2D}}T_{\mu\nu\rho\sigma}(|x_1^\delta-x_2^\delta|)
\end{equation*}
where $T_{\mu\nu\rho\sigma}$ is a combination of the $I_{\mu\nu}$ tensors, given in equation (\ref{itensor}). This simply keeps track of the transformation of the Lorentz indices. Notice the constant $c$ in the numerator is not just any constant. Rather, it is a very important constant, known as the \textit{central charge}. This number will be discussed in the next section.

\subsection{Central Charge}

The Central Charge (usually denoted by a lower case, $c$) is a very important quantity in any CFT. It is one of the quantities defining what specific CFT you are studying. It is roughly a measure of the the number of degrees of freedom of your system. You may recall from your Statistical Mechanics course that the number of degrees of freedom of a system is associated with how many pieces of information you need to describe the state of your system. For instance, the number of position degrees of freedom of a point particle in a 3D box is three (one for each coordinate). If you constrain your particle to move on a plane, then you are left with only two position degrees of freedom. Alternatively, you could replace your point particle with a molecule. Now, you may need to add degrees of freedom associated with the orientation of your molecule. In a QFT, the central charge is roughly a measure of this. 

There are some very important theorems that come up involving the central charge. The first is known as the c-theorem, which states that there exists a function $C(g_i,\mu)$ (where $g_i$ are the couplings and $\mu$ is the energy scale) in QFT that:
\begin{itemize}
    \item decreases monotonically\footnote{This means it only decreases or remains constant, it does not increase at any point.} under RG flow.
    \item at fixed points (where CFTs live) it becomes a constant.
\end{itemize}
Alexander Zamolodchikov proved that such a function always exists in 2D CFTs \cite{Zamolodchikov:1986gt} and that at the fixed points it is equal to the central charge. There has been work on defining such a function in higher dimensions. In four dimensions, it is referred to as an a-theorem. This is thanks to John Cardy, who conjectured that a certain anomaly coefficient would do the trick (thus the name) \cite{CARDY1988749}. Hugh Osborne and Ian Jack first proved this perturbatively \cite{Jack:1990eb} and later, Komargodski and Schwimmer proposed a non-perturbative proof.

\subsection{Ward Identities}

In Quantum Field Theory, one learns that the Ward Identities are a set of differential equations that relate correlation functions with insertions of conserved currents to those without. Therefore, given an $n$-point correlator, one can solve a differential equation to find an $(n+1)$-point correlator, where the additional operator is a conserved current. These are a powerful tool in computing correlation functions.

We learned that the stress tensor is conserved. In CFT, applying the Ward identities with an insertion of the stress tensor can be very powerful. In this case, the differential equation is given by
\begin{equation}
    \partial_\mu\langle T^{\mu\nu}(x)\hat{\mathcal{O}}_1(x_1)\dots\hat{\mathcal{O}}_n(x_n)\rangle=-\sum_{i=1}^n\delta(x-x_i)\langle \hat{\mathcal{O}}_1(x_1)\dots\hat{\mathcal{O}}_n(x_n)\rangle
\end{equation}
As you can see, if you know the $n$-point function on the right-hand-side, you can substitute it in and then are left with a differential equation for the $(n+1)$-point correlator with the stress tensor inserted.

\subsection{Radial Quantization}

When quantizing a particular theory, generally one must complete the following steps:

\begin{itemize}
    \item[] \textbf{Step 1.} Foliate spacetime into fixed ``time" slices.
    \item[] \textbf{Step 2.} Define states on each time slice.
    \item[] \textbf{Step 3.} Evolve these states between the time slices by operating on them with the Hamiltonian.
\end{itemize}

From relativity, we know that there is no one way to pick your time slicing. Any reference frame is as good as any other. One can use this freedom strategically in order to make things simpler. For instance, it is ideal if your fixed time slices are chosen so that they respect the symmetries of your theory. If you do this then the Hilbert space that lives on each slice will be the same because they are related by some symmetry transformation.

When working in a Lorenzian signature, you can choose a time-like coordinate and your fixed time-slices are at fixed values of that coordinate.  On the other hand, in CFT, we often deal with Euclidean space, where there is no time dimension. This is why we put time in quotes in the steps above. If there is no time, then how can you foliate into time-slices? Well, you can't. However, you can still foliate your space in some sort of slicing. One might be inclined to just pick one of your Cartesian coordinates and foliate based on fixed values of that coordinate. However, which one should you choose? Each Cartesian direction should be as good as any other. It turns out that for CFTs, a good choice is to work in spherical coordinates. In that case, you have only one radial dimension. Therefore, this is a natural coordinate to single out. Each ``time-slice" is therefore living on one of infinitely many concentric spheres, one at each fixed radius. This works because CFTs have scale-invariance. You transform between each sphere by acting with the dilitation operator. Therefore, your foliation still respects the symmetries of your theory. This process in called \textit{radial quantization}.

In radial quantizaion, you prepare your initial state on some sphere and then evolve the state by acting on it with the dilitation operator. This is analogous to preparing your initial state at some fixed time (often $t=0$) and then evolving it forward in time with the Hamiltonian. In radial quantization, our dilitation operator therefore acts like a kind of Hamiltonian.

\subsection{State-Operator Correspondence} \label{stateoperator}

In CFTs, there is a one-to-one correspondence between states and operators acting at the origin. To better understand this, let's consider working in our Dilitation Operator Eigenbasis. We generate eignstates of $\hat D$ by acting with operators with definite dimension at the origin $\hat{\mathcal{O}}_\Delta(0)|0\rangle=|\Delta\rangle$. You might wonder what will happen if you act with an operator at a point other than the origin. It turns out that this state will not be an eigenstate of $\hat D$, but instead can be written as a superposition of eigenstates of $\hat D$. Therefore, even general states (not necessarily only eigenstates of $\hat D$) can be generated by considering a superposition of states that were created by acting with an operator at the origin. Furthermore, we can go the other way. For every state, we can find an operator acting at the origin. 

\subsection{Operator Product Expansion} \label{OPE}

In QFT, the operator product expansion (OPE) says that you should be able to replace two local operators by a series of operators inserted at the midpoint between the two, in the limit that the operators become close together. Hopefully this sounds reasonable. You imagine starting with two operators, you bring them near each other, and then zoom out until you can no longer tell that it is two operators. You can then represent the two by a linear superposition of single operators. This is helpful in QFT, but is especially powerful in CFT because of scale invariance.

To illustrate this explicitly, imagine you have two operators, $\hat{\mathcal{O}}_1(0)$ and $\hat{\mathcal{O}}_2(x)$, acting at the origin and at some arbitrary point $x$, respectively. This will produce some state, $\ket{\psi}$, which lives on a sphere centered at the origin and containing the point $x$ (remember we are working in radial quantization, where our states live on spheres).
\begin{equation*}
\ket{\psi}=\hat{\mathcal{O}}_2(x)\hat{\mathcal{O}}_1(0)\ket{0} 
\end{equation*}
We also know via the state operator correspondence, that any general state can be found by taking a superposition of states created by acting with operators at the origin.
\begin{equation}
\ket{\psi}=\sum_nc_n\mathcal{O}_\Delta(0)\ket{0}=\sum_nc_n\ket{\Delta_n}
\end{equation}
where $\ket{\Delta_n}$ is generated by acting on the origin with primaries or descendants. Combining these two facts, we are left with $$\hat{\mathcal{O}}_2(x)\hat{\mathcal{O}}_1(0)\ket{0}=\sum_{\mathrm{primaries}}C_\mathcal{O}(x,\partial_y)\mathcal{O}(y)\big |_{y=0}\ket{0}$$
where we have re-expressed our descendants as derivatives acting on primaries. We see that the two operators on the left-hand-side can be rewritten as a sum over single operators that act at the origin.

We know this will converge whenever we are able to draw a sphere around the two operator insertions of interest that does not enclose any other operators. This convergence is what makes the OPE so powerful in CFT.

\subsection{Conformal Bootstrap} \label{bootstrap}
The Conformal Bootrap Program is a non-Lagrangian, axiomatic approach to solving CFTs exactly. In standard QFT, you take the Lagrangian to be the defining object for your system. The Conformal Bootstrap Program takes an entirely different approach. Instead, it takes the dimensions or your operators and OPE coefficients as the defining information. A key strategy that is used in this program is to enforce the Operator Product Expansion that was introduced in the previous section. This allows you to place additional constraints on, for instance, your four-point functions. These methods have been incredibly fruitful for CFTs in dimensions $D\geq 3$.

\subsection{Embedding Space Formalism}
The Conformal Group in $D$ dimensions is given by $SO(D,1)$. But, wait? You might recall that the connected part of the Lorentz Group was described by the Special Orthogonal Group. Notice, that the number of dimensions is different. $SO(D,1)$ is the Special Orthogonal Group in $D+1$ dimensions. What we learn then is that the Conformal Group can be described by the Lorentz Group on one higher dimension. This fact is the foundation of the \textit{Embedding Space} Formalism. This is a method for studying CFTs in $D$ dimensions by studying a Lorentz invariant theory in $(D+1)$-dimensions. You are embedding your system in a higher dimensional theory. This allows you to use potentially more familiar tools in the higher dimensional space and then you simply need to project down to the lower dimensional theory at the end of the calculation. For more on this, see, e.g., \cite{dirac,Weinberg_2010,Costa_2011}.

\subsection{When does Scale Invariance Imply Special Conformal Invariance?}
Historically, in studying different examples of CFTs, researchers began to notice something interesting, which was that whenever a Lorentz invariant theory was scale invariant (i.e. invariant under dilitations), then it was ultimately invariant under the full conformal group. That is, if the theory was invariant under a uniform scale transformation (where $\Omega(x)=$ constant), it was also invariant under all angle-preserving, position-dependant re-scalings. Theoretically, it is possible for a theory to be invariant under only the subgroup of Lorentz plus dilidations. One might wonder then if somehow scale invariance implies conformal invariance? Is there anything that needs to additionally enforced for this to be true? Is this true in all numbers of dimensions (e.g. is it special to 2D?)? 

In \cite{Zamolodchikov:1986gt} and \cite{Polchinski:1987dy}, the authors were able to prove that if one enforces unitarity in addition to scale invariance, then one will always end up with a theory that is invariant under the full conformal group in two-dimensions.  There is no proof of this in higher dimensions, however it is believed that scale plus unitarity are enough to get conformal invariance in $D\leq6$.

\section{Summary}
In this primer, we introduced many of the key concepts and methodologies in Conformal Field Theory (CFT) in $D\geq3$. We learned that a CFT is a field theory with conformal symmetry.  This means that physical quantities, like correlation functions, should be invariant under translations, rotations/boosts, dilitations, and the special conformal transformation. In this primer, we dealt primarily with quantum CFTs, though one can also consider classical CFTs.

One of the key takeaways we would like all of the readers to leave with is that symmetry is a very powerful tool. We discussed the fact that a (quantum) CFT is simply a QFT with more symmetry. By enforcing invariance under all symmetry transformations, we were able to derive the spatial dependence of one-, two-, and three-point functions. This means that this is a universal feature of all CFTs, no matter how different the physical systems may appear.

Finally, CFTs are important for the following reasons and more. 
\begin{itemize}
    \item They help us to understand the full space of QFTs.
    \item We can perturb away from a CFT to learn about nearby QFTs.
    \item Many statistical and condensed matter systems are described by a CFT.
    \item String Theory, which is a candidate framework for Quantum Gravity, is a 2D CFT
    \item The AdS/CFT correspondence and other gauge/gravity dualities are tools that allow us to learn about gravitational systems by performing CFT calculations. 
\end{itemize}

\section*{Acknowledgements}
This work was supported by the Department of Energy under grant {DE-SC} 0023876; the Bryant and Diane Hichwa Research Award; the Horace L. Newkirk Research Award; and the SSU RSCAP Summer Fellowship.

\appendix
\section{The Jacobian}
    \label{jacobian}
The Jacobian is the determinant of the Jacobian matrix. The Jacobian matrix is a matrix that carries information about the way a function changes when you make a change of coordinates. In particular, if you start with some coordinates $x^\mu$ and then transform to another set of coordinates $x^\mu\rightarrow \tilde x^\mu (x)$, the Jacobian matrix is given by
\begin{equation}
    J_{\mu \nu}=\frac{\partial \tilde x^\mu}{\partial x^\nu} 
\end{equation}
where $J_{\mu\nu}$ is the Jacobian matrix. 

To see a familiar example, consider transforming from Cartesian ($x,y,z$) coordinates to Spherical ($r,\theta,\phi$) coordinates. 
\begin{equation*}
   x\hat{x}+y\hat{y}+z\hat{z} = r\sin\left( \theta \right)\cos\left( \phi \right)\hat{x} +  r\sin\left( \theta \right) \sin\left( \phi \right)\hat{y} +  r\cos\left( \theta \right)\hat{z}  
\end{equation*}
Our Jacobian Matrix is given by
\begin{align*}
    J_{\mu \nu} &=
    \begin{bmatrix}
        \partial_{1} f_1 & \partial_{2} f_1 & \partial_{3} f_1\\
        \partial_{1} f_2 & \partial_{2} f_2 & \partial_{3} f_2\\
        \partial_{1} f_3 & \partial_{2} f_3 & \partial_{3} f_3\\
    \end{bmatrix}\\
    &=
    \begin{bmatrix}
        \partial_{r} r\sin\left( \theta \right)\cos\left( \phi \right) & \partial_{\theta} r\sin\left( \theta \right)\cos\left( \phi \right) & \partial_{\phi} r\sin\left( \theta \right)\cos\left( \phi \right)\\
        \partial_{r} r\sin\left( \theta \right) \sin\left( \phi \right) & \partial_{\theta} r\sin\left( \theta \right) \sin\left( \phi \right) & \partial_{\phi} r\sin\left( \theta \right) \sin\left( \phi \right)\\
        \partial_{r} r\cos\left( \theta \right) & \partial_{\theta} r\cos\left( \theta \right) & \partial_{\phi} r\cos\left( \theta \right)\\
    \end{bmatrix}\\
    &=
    \begin{bmatrix}
        \sin\left( \theta \right)\cos\left( \phi \right) & r\cos\left( \theta \right)\cos\left( \phi \right) & -r\sin\left( \theta \right)\sin\left( \phi \right)\\
        \sin\left( \theta \right) \sin\left( \phi \right) & r\cos\left( \theta \right) \sin\left( \phi \right) & r\sin\left( \theta \right) \cos\left( \phi \right)\\
        \cos\left( \theta \right) & -r\sin\left( \theta \right) & 0\\
    \end{bmatrix}\\
\end{align*}
To find the Jacobian ($J$) we take the determinant of $J_{\mu \nu}$
\begin{align*}
    J &=
    \cos\left(\theta\right)
    \begin{vmatrix}
        r\cos\left(\theta\right)\cos\left(\phi\right) & -r\sin\left(\theta\right)\sin\left(\phi\right) \\
        r\cos\left(\theta\right)\sin\left(\phi\right) & r\sin\left(\theta\right)\cos\left(\phi\right) 
    \end{vmatrix}
    +r\sin\left(\theta\right)
    \begin{vmatrix}
        \sin\left(\theta\right)\cos\left(\phi\right) & -r\sin\left(\theta\right)\sin\left(\phi\right) \\
        \sin\left(\theta\right)\sin\left(\phi\right) & r\sin\left(\theta\right)\cos\left(\phi\right)
    \end{vmatrix}\\    
    &= r^2\sin\left(\theta\right)
\end{align*}
The Jacobian is hopfully familiar from your multivariable calculus class. There you learned that when integrating in spherical coordinates the differential $dr d\theta d\phi$ is multiplied by a factor of exactly $r^2\sin\left(\theta\right)$. Explicitly, 
\begin{align*}
\int_V f\left(x,y,z\right) \,dxdydz &= \int_V f\left(r, \theta, \phi\right) \,r^2\sin\left(\theta\right) dr d\theta d\phi\\
\end{align*}
This is true for any change of coordiantes when you are performing a multivariable integral.
\begin{align*}
\int f\left(x_1,...,x_n\right) \,dx_1...dx_n &= \int f\left(x^\prime_1,...,x^\prime_m\right) \,J dx^\prime_1...dx^\prime_m\\
\end{align*}
Where $J$ is given by 
\[J = \det
    \begin{vmatrix}
        \partial_{x^\prime_1} f_1 & ... & \partial_{x^\prime_n} f_1 \\
        \vdots & \ddots & \vdots \\
        \partial_{x^\prime_1} f_m & ... & \partial_{x^\prime_n} f_m \\
    \end{vmatrix}
\]
Conceptually the Jacobian can be thought of as a scale factor that gives information on what happens to the volume of our differential in a coordinate system under a coordinate transformation. If we consider the transformation between 2D Cartesian and polar coordinates\footnote{This is a good exercise for practicing use of the Jacobian} we find $J = r$. This tells us that the box $dxdy$ is replaced by $rdrd\theta$. The area of the differential is now a function of the radius. The ``boxes'' on a polar coordinate chart get bigger the farther they are from the origin.

\section{Quantum Harmonic Oscillator} \label{HarmOsc}
Recall the Simple Harmonic Oscillator Potential
\[V(x) = \frac{1}{2} m \omega^2 x^2\]
This is a conservative system and our Hamiltonian is simply given by kinetic plus potential energy. 
\begin{align*}
    H &= T + V(x) \\
    H &= \frac{1}{2}mv^2 + \frac{1}{2} m \omega^2 x^2 \\
    H &= \frac{p_x^2}{2m} + \frac{1}{2} m \omega^2 x^2 \\
\end{align*}
When we quantize our system, our variables become operators. In the Quantum Harmonic Oscillator,  our Hamiltonian operator is given by
\begin{align*}
    \hat{H} &= \frac{\hat{p}_x^2}{2m} + \frac{1}{2} m \omega^2 \hat{x}^2 \\
\end{align*}
We want to solve for our energy eigenstates. There is a neat trick for doing this by defining new operators known as the raising operator, the lowering operator, and the number operator. The naming scheme of these operators will soon become apparent.
\begin{align*}
    \text{Raising operator: }\hat{a}^\dagger &= \sqrt{\frac{m\omega}{2\hbar}}\left(\hat{x} - \frac{i}{m\omega}\hat{p}_x\right) \\
    \text{Lowering operator: }\hat{a} &= \sqrt{\frac{m\omega}{2\hbar}}\left(\hat{x} + \frac{i}{m\omega}\hat{p}_x\right) \\
    \text{Number Operator: } \hat{N} &= \hat{a}^\dagger\hat{a}
\end{align*}
It is left to the reader to show the following for these new operators
\begin{enumerate}
    \item $\left[ \hat{a}, \hat{a}^\dagger\right] = 1$
    \item $\hat{x} = \sqrt{\frac{\hbar}{2\omega m }}\left(\hat{a} + \hat{a}^\dagger \right)$
    \item $\hat{p}_x = -i\sqrt{\frac{2\omega m }{\hbar}}\left(\hat{a} - \hat{a}^\dagger \right)$
    \item $\hat{H} = \hbar \omega \left( \hat{a}^\dagger\hat{a} + \frac{1}{2}\right)=\hbar \omega \left( \hat{N} + \frac{1}{2}\right)$
    \item $[\hat H, \hat N]=0$
    \item $\left[ \hat{N}, \hat{a}^\dagger\right] = \left[ \hat{H}, \hat{a}^\dagger\right]= \hat{a}^\dagger$
    \item $\left[ \hat{N}, \hat{a} \right] = \left[ \hat{H}, \hat{a} \right] = -\hat{a}$
\end{enumerate}
The fact that the number operator $\hat N$ commutes with the Hamiltonian $\hat H$ means that the two have a simultaneous set of eigenstates. That is, the energy eigenstates are also eigenstates of the number operator. We therefore will label our energy eigenstates as $|n\rangle$, where the $n$ refers to the eigenvalue of the state associated with the number operator.

By enforcing the commutation relations in (6) and (7), one can show that the state given by acting with the raising operator $\hat a^\dagger$ on an energy eigenstate increases the value of $n$ by one, i.e. $\hat a^\dagger |n\rangle=|n+1\rangle$. The lowering operator $\hat a$, on the other hand, decreases the eigenvalue by one, i.e. $\hat a |n\rangle=|n-1\rangle$

This means, we can generate the full list of energy eigenstates by acting with raising and lowering operators. We define the \textit{lowest weight state} to be the state that has $n=0$ and is therefore annihilated by the lowering operator.

\section{Quick Reference Dictionary} \label{dictionary}

This Dictionary includes very short definitions of some of the key terms in CFT. In line with the rest of this paper, rather than providing technical definitions, these are more conceptual descriptions. Each definition includes a reference to the section in the primer where the term is discussed in more detail.

\vspace{0.5cm}

\begin{dict}{Beta Function, \ref{rgflow}}
    This function describes how coupling constants in a QFT change as you change the energy scale. CFTs live where beta functions vanish and therefore are at critical points along the RG Flow.
\end{dict}
\begin{dict}{Commutator, \ref{QM}} 
    Quantification of the amount by which two operators fail to commute, $[\hat{\mathcal{O}}_1,\hat{\mathcal{O}_2}]=\hat{\mathcal{O}_1}\hat{\mathcal{O}_2}-\hat{\mathcal{O}_2}\hat{\mathcal{O}_1}$. Operators that commute have simultaneous eigenstates.
\end{dict}
\begin{dict}{Conformal Family, \ref{Primaries}} 
    A group of operators consisting of one primary operator along with all of its descendants. In a CFT, all operators can be organized into Conformal Families.
\end{dict}
\begin{dict}{Conformal Transformations, \ref{confsymm}} 
    The set of transformations containing translation, rotation, dilatation, and the special conformal transformation.
\end{dict}
\begin{dict}{Correlation Function, \ref{correlationfunct} and \ref{correlators}}
    A measure of how much different local operators acting at different positions are correlated.
\end{dict}
\begin{dict}{Descendants, \ref{Primaries}} 
    All operators in your CFT can be classified as Primary or Descendant Operators. Descendants are created by acting on a Primaries with the raising operator, $\hat P_\mu$, some number of times.  
\end{dict}
\begin{dict}{Euclidean Signature and Euclidean Space, \ref{SR}}
    Here, you treat all of your dimensions as spatial dimensions (there is no time dimension). The Minkowski metric in Euclidean signature as all positives along the diagonal. Much of the CFT literature is in Euclidean signature.
\end{dict}
\begin{dict}{Expectation Value, \ref{QM}}
    The average value of infinitely many identical measurements made on infinitely many identically prepared states.
\end{dict}
\begin{dict}{Field, \ref{QFT}} 
    A mathematical object that takes position as its input. You can have different types of fields that return different types of mathematical objects. For instance, a scalar field returns a scalar, while a vector field returns a vector.
\end{dict}
\begin{dict}{Generators, \ref{lie}} 
    Generators are infinitesimal transformations that can be used to build up larger transformations. These are the elements of your Lie Algebra.
\end{dict}
\begin{dict}{Hermitian Operator, \ref{QM}}
    An operator such that $\hat{H}^\dagger=\hat{H}$. In Quantum Mechanics, observables are represented by Hermitian Operators.
\end{dict}
\begin{dict}{Irrelevant Operator, \ref{classification}} 
    An operator with dimension greater than the number of spacetime dimensions being considered, $\Delta>D$. These become less important as you flow to lower energies.
\end{dict}
\begin{dict}{Local Operators, \ref{QFT}} 
    Operators that are a function of position, e.g. $\hat{\mathcal{O}}(x^\mu)$.
\end{dict}
\begin{dict}{Lorentzian Signature, \ref{SR}} 
    This described a spacetime with one time dimension. The Minkowski Metric in Lorentzian signature as one negative sign.
\end{dict}
\begin{dict}{Lorentz Transformations, \ref{poincare}} 
    The set of transformations containing rotations and boosts.
\end{dict}
\begin{dict}{Lowest Weight State, \ref{Primaries} and \ref{HarmOsc}} 
    This term comes from the approach to solving the Simple Harmonic Oscillator with raising and lowering operators. The lowest weight state is the one that is annihilated by the lowering operator.
\end{dict}
\begin{dict}{Marginal Operator, \ref{classification}} 
    An operator with dimension equal to the number of spacetime dimensions being considered, $\Delta=D$.
\end{dict}
\begin{dict}{Mass Dimension, \ref{dimension}}
    Work in units with $c=\hbar=1$. Then, using dimensional analysis, you can determine the dimension of objects in terms of powers of mass.
\end{dict}
\begin{dict}{Metric, \ref{SR}} 
    A mathematical object that describes the geometry of your spacetime in General Relativity.
\end{dict}
\begin{dict}{Minkowski Space, \ref{SR}} 
    This is the geometry of Special Relativity, when your spacetime is flat.
\end{dict}
\begin{dict}{OPE Coefficients, \ref{threepoint} and \ref{OPE}} 
    The Coefficients that appear in the Operator Product Expansion. These are equal to the Three-Point coefficients, so these names are often used interchangeably.
\end{dict}
\begin{dict}{Operator Dimension $\Delta$, \ref{QFT}} 
    All operators have an associated dimension. This specifies how the operators will transform under spacetime dilatations. 
\end{dict}
\begin{dict}{Orthochronous Transformations, \ref{poincare}}
    These are transformations that preserve the direction of time.
\end{dict}
\begin{dict}{Orthogonal Group $O(D)$, \ref{poincare}}
    The set of matrices $S=\{\Lambda_1,\Lambda_2,\Lambda_3,...\}$ such that $\Lambda^T\Lambda=I$.
\end{dict}
\begin{dict}{Poincar\'e Transformations, \ref{SR}} 
    The set of transformations containing Lorentz transformations and translations.
\end{dict}
\begin{dict}{Primary Operator, \ref{Primaries}} 
    All operators in your CFT are either Primaries are Descendants. These operators are the ``lowest weight" operators in your CFT. By definition, these commute with the SCT operator, $[\hat{\mathcal{O}}(0),\hat K_\mu]=0$. If you act on the vacuum with a Primary, the resultant state is annihilated by $\hat K_\mu$. Additionally, these operators have nice transformation properties. They transform as tensor densities.
\end{dict}
\begin{dict}{Relevant Operator, \ref{classification}}
    An operator with dimension less than the number of spacetime dimensions being considered, $\Delta<D$. These become more important as you flow to lower energies.
\end{dict}
\begin{dict}{Scaling Dimension $\Delta$, \ref{dimension}}
    Information regarding how an operator will transform under dilatation. This is another term for Operator Dimension.
\end{dict}
\begin{dict}{Special Orthogonal Group $SO(D)$, \ref{poincare}} 
    The set of Lorentz transformations, $\Lambda$, such that $det(\Lambda)=+1$. These transformations preserve orientation.
\end{dict}
\begin{dict}{Structure Constants, \ref{threepoint}} 
    Another name for Three-point Coefficients or OPE Coefficients.
\end{dict}
\begin{dict}{Three-Point Coefficients, \ref{threepoint}}
    The constant coefficients that appear in the three-point function for a particular set of three operators. This is going to be determined by how strongly the operators interact. 
\end{dict}
\begin{dict}{Unitary Operator, \ref{QM}}
    An operator such that $\hat{U}^\dagger \hat{U} = \hat{I}$. In Unitary Quantum Mechanics, time evolution is governed by a Unitary Operator. This allows the standard probabilistic interpretation of QM to make sense, as it preserves the normalization of states.
\end{dict}

\bibliographystyle{unsrt}
\bibliography{refs}

\end{document}